\documentclass[hyper]{JHEP3}

\newcommand{\guuu}{\raisebox{2pt}{$\bar{\hspace{9.5pt}}$}\hspace{-9pt}\raisebox{3.5pt}{$\bar{\hspace{9.5pt}}$}\hspace{-10.25pt}
\bar{G}}
\newcommand{\quu}{\raisebox{2pt}{$\bar{\hspace{10pt}}$}\hspace{-10pt}\bar{Q}}
\newcommand{\quuu}{\raisebox{2pt}{$\bar{\hspace{9.5pt}}$}\hspace{-9pt}\raisebox{3.5pt}{$\bar{\hspace{9.5pt}}$}\hspace{-10.25pt}
\bar{Q}}
\newcommand{\qdd}{Q\put(-9.25,-2){$\line(1,0){6}$}\put(-9,-1){$\line(1,0){6}$}}
\newcommand{\qddd}{Q\put(-9.5,-3){$\line(1,0){6}$}\put(-9.25,-2){$\line(1,0){6}$}\put(-9,-1){$\line(1,0){6}$}}
\newcommand{\n}{\nonumber}
\usepackage{graphicx}

\title{On the Moduli Space of SU(3) Seiberg-Witten Theory with
Matter}
\author{Brett J. Taylor \\ \\
Department of Physics \\ University of Patras \\ Achaia 26500\\
Greece\\ E-mail: \email{btaylor@pythagoras.physics.upatras.gr}}
\abstract{ We present a qualitative model of the Coulomb branch of
the moduli space of low-energy effective ${\cal N}=2$ SQCD with
gauge group SU$(3)$ and up to five flavours of massive matter.
Overall, away from double cores, we find a situation broadly
similar to the case with no matter, but with additional complexity
due to the proliferation of extra BPS states. We also include a
revised version of the pure SU$(3)$ model which can accommodate
just the orthodox weak coupling spectrum.}


\preprint{}

\begin{document}

\parindent5pt
\parskip5pt
\pagebreak
\section{Introduction}

There was much excitement caused by the seminal papers of Seiberg
and Witten \cite{sw1,sw2} in which they found exact
non-perturbative results for low-energy effective ${\cal N}=2$
supersymmetric gauge theories. A wide array of authors have
expanded upon the initial work
\cite{bf1,bf2,bf3,kl1,kl2,klt,ad1,bl,dsund}, both in fleshing out
further detail within the SU$(2)$ case, and also boldly entering
into the domain of the higher gauge groups. It is no less apparent
now than originally that these results are more than those of a
toy model, but rather present an iconic paradigm, a nugget of
solidity up against which most new abstract theories must be
measured. It has already been generalised and deformed in a myriad
of ways, and we expect this to continue in the years to come. We
consider it important, therefore, to work through some of the
remaining details in the more complex, but still amenable cases.
In particular, the example of SU$(3)$ is one which merits
attention, being both the simplest case where the two Higgs
scalars of the theory generically misalign in group space; and, of
course, due to the links with the usual conundrum that is QCD.

In \cite{me} we presented a plausible, consistent qualitative
model of the curves of marginal stability that exist in the moduli
space of pure SU$(3)$ Seiberg-Witten Theory. In addition to
modifying this slightly, we shall now show how this generalises to
the cases with up to five matter multiplets in the fundamental
representation of the gauge group.

In each case we have a four real-dimensional space which we
choose, following \cite{ad1}, to write as a fibration of
concentric three-spheres. There exists a surface of points on
which this abelian theory breaks down due to extra fields becoming
massless. The intersection of this with a three-sphere with radius
larger than a critical value comes in the shape of a knot called a
trefoil. In the pure case there is an instanton effect which
exacts a bifurcation of the classical trefoil. For $N_{f}$
flavours, this local bifurcation patches together non-trivially -
there are $N_{f}$ half-twists as we traverse the classical knot
(so for odd $N_{f}$ we retain one, more complicated, knot). In
addition there are in general $N_{f}$ singular curves
corresponding to where the quarks (with non-degenerate bare
masses) become massless. These are generally planes, which
therefore intersect our three-sphere as circles, until we increase
the bare masses sufficiently whereupon they shrink to point before
giving a null intersection.

At a point of weak coupling, we have the BPS spectrum first given
in \cite{dnt}. This consists of massless gauge bosons plus the
same towers of dyon hypermultiplets as in the pure case; the quark
hypermultiplets, often with bare mass; and massive gauge bosons
which can be thought of as bound states of two of these. Finally,
the largest set of states are bound states of dyons and quarks.
Each dyon can bind in up to $2^{N_{f}}$ combinations.

As we increase the coupling, there exist hypersurfaces past which
some BPS bound states are no longer stable. In a region of
moderate to strong coupling, we see most of the dyon-quark states
lose stability. Curiously, the flavour singlet states consisting
of a dyon together with one of each flavour quark remain stable
into the regions of strong coupling in root directions not equal
to the magnetic charge of that dyon. This is true subject to the
proviso that the dyon is stable in that region. In analogy with
the pure SU$(3)$ example, there are single cores, where one
abelian gauge symmetry is strongly coupled; and double cores where
both are. In the single core, one of the towers of dyons, except
for two of its elements, loses stability; and the other two towers
are transformed. In the double core, most of these others become
unstable too, and we are left with a finite spectrum. We have such
cores in all the examples. The only addition to the spectrum now
is that of all the quarks (and some states into which they can
transform when passing through a branch cut), plus the flavour
singlet bound states with dyons of the reduced spectrum.

We have produced this somewhat sketchy atlas as a first attempt to
gain a concrete understanding of these moduli spaces. Hopefully
this will inspire a more rigorous study by the more determined or
enlightened. Our findings contain nothing startlingly contentious
and are self-consistent. Our weak coupling spectrum is orthodox,
following \cite{lee}. Our strong-coupling spectrum agrees with
that of \cite{dnt} at the point where they show it must be dual to
that of the $\mathbb{CP}^{5-N_{f}}$ sigma model, for
$6>N_{f}\geq3$. We present what we believe lies in between.
However, we feel we should point out before proceeding that this
study is but a tower of logic crying out for the firm basis a plot
of these curves would provide.

Thus we begin in Chapter two with a review of the pure case, in
which we include the slight modifications to SU$(3)$ SYM. In
Chapter three we sketch the SU$(2)$ picture in the same manner as
for SU$(3)$ in order to allow for comparison, then we consider in
detail the case with one flavour. In Chapter four we deal mainly
with the example of three flavours to show the pattern followed by
all the rest. Chapter five looks at what must happen as we flow
between the various examples. We conclude with a brief discussion
of how we agree with other work.

\section{The pure case revisited}

\subsection{Classical Model}

Let us begin with a review of the pure SU$(3)$ model, further
details can be found in \cite{me}. SU$(3)$ is a group based on an
eight-dimensional Lie algebra, within which we find embedded in
the standard way three copies of SU$(2)$. We think of these as
lying along the root directions of the Cartan subalgebra (CSA),
which are harmoniously arranged within a plane to give hexagonal
symmetry. An SU$(N)$, ${\cal N}=2$ gauge theory has two adjoint
Higgs fields in its gauge multiplet. We consider only the Coulomb
branch, at a generic point of which these spontaneously break the
gauge symmetry to its maximal torus. The vacuum expectation values
of these determine the parameters of each theory. We range over
them as we seek to determine the moduli space of supersymmetric
vacua, and at each point also we require the spectrum of excited
states that are BPS and so amenable to us of the second
superstring revolution. Gauge-fixing transformations allow us the
freedom to conjugate these vevs into the maximal torus, so we can
restrict ourselves to the (complexified) CSA (then exponentiate).
Even so, generically the two Higgs fields do not align. If either
lies perpendicular to a root direction we have trouble. Dark
(non-abelian) clouds loom if our adjoint Higgs do not induce
spontaneous symmetry breaking of SU$(3)$ down to U$(1)\times$
U$(1)$. Seiberg and Witten \cite{sw1} showed that this manifests
itself at the boundary of the closure of the moduli space where a
field that we have considered massive in the effective theory
becomes massless, conveying the effect of the symmetry
restoration. We call this set the singularity curve (though our
sections of it may make it seem disconnected into many curves). It
can be determined as the zero set of a high-order polynomial in
the coordinates we choose for the moduli space. As a surface it
has singularities in the mathematical sense. Thankfully, for
SU$(3)$ these are only points. Particularly interesting conformal
field theories often lie here, however \cite{ad1}, and the areas
of the moduli space with a finite stable BPS spectrum, nearby.

The complexified CSA of SU$(3)$ has four real dimensions, which we
parameterise by two complex variables, $u$ and $v$, the quadratic
and cubic Casimirs of the algebra, natural as these are Weyl group
invariant, a symmetry left over after gauge fixing to the CSA.
They have mass dimensions of two and three respectively. All the
classes of theories have a dynamically generated scale $\Lambda$
(although it depends on $N_{f}$ and their masses). Setting this to
zero eliminates the non-perturbative effects and is what we mean
by the `classical case' (1-loop perturbative would be more
accurate). Here the singularity curve is given by
$4u^{3}=27v^{2}$. Following \cite{ad1} we find the intersection
with the (topological) 3-sphere defined as
$4|u|^{3}+27|v|^{2}=R^{6}$. This means that $u$ and $v$ are of
constant length and cycle (on a torus) in the ratio $3:2$.

\FIGURE[h]{ \centering
\includegraphics[width=5cm]{trefcl.pstex}
\caption{The classical trefoil ${\cal T}_{cl}$ wrapping three
times around one cycle as it wraps twice around the other.}}

Physical quantities are obtained not from $u$ and $v$, but from
appropriate linear combinations of the complex 2-vectors $a$ and
$a_{D}$, where $a$ is the vev of the Higgs. These are more
accurately sections of an Sp$(4,\mathbb{Z})$ bundle over the
moduli space, and they show up in the central charge of the
algebra, which relates to the energy of BPS states.  The rest mass
of a BPS state $(g,q)$ with magnetic charge $g$ and electric
charge $q$ is given by
\begin{equation}
m[ (g, q)] = |g.a_{D}+ q.a|.
\end{equation}
Upon circling any hole in the moduli space, $(a, a_{D})$ is acted
upon by an element of Sp$(4,\mathbb{Z})$. Obviously $(g, q)$ must
be acted upon by the inverse element of Sp$(4,\mathbb{Z})$ to
maintain the mass. Pick a basis $\{ (g_{1}, q_{1}),\dots
,(g_{n},q_{n}) \} $ of states that become massless at the
singularity we are circling, then the transformation matrix is
singled out as a product of ones of the canonical form for each
element
\[
M_{(g,q)}= \left(
\begin{array}{c|c}
1+q\otimes g & q \otimes q \\ \hline -g \otimes g &
\mbox{\hspace{5pt}}1-g\otimes q
\end{array}
\right).
\]

We are free to choose where to enact this transformation on the
charges (and $a$ and $a_{D}$), but somewhere we must decide to put
a branch cut (which is one real-codimensional) emanating from the
(one complex-codimensional) singular curve. We must also decide
where we would like it to go. For simplicity we shall make most of
our cuts meet at the origin.

As well as gauge bosons, we learn from \cite{tim1, tim2,timchrist,
lee} among others that the weak-coupling spectrum includes dyons
in massive ${\cal N}=2$ hypermultiplets. The magnetic charge of
each dyon must be root-valued, {\it i.e.}\ has charge $\alpha_{1},
\alpha_{2}$ or $\alpha_{3}$ (if it is negative, its CP conjugate,
which will also be present will be positive, so we take this).
These occur in towers, with one of each root coexisting at each
point. The tower of the non-simple root out of these (defined as
the sum of the two obtuse roots having positive scalar product
with the dominant principal direction of the Higgs fields) is
comprised of stable bound states of the other two and is slightly
different. The other, more basic two, are simply embeddings of the
SU$(2)$ dyons in the simple root directions and as such have
electric charge in integer multiples of their magnetic charge. The
electric charge of the third tower is `offset'. Each state is a
bound state of $n$ units of one root with either $n+1$ or $n-1$ of
the other. These two bound states do not co-exist and hence form
two distinct weak-coupling limits (in our picture, these
correspond to the origin and infinity). For definiteness, we shall
concentrate on the third of the moduli space for which
$\alpha_{3}$ is non-simple. This has regions, among others,
extending from the two limits containing the states
\[ Q_{1}^{n}= (\alpha_{1},(n-1)\alpha_{1}) \hspace{5pt},\hspace{5pt}
Q_{2}^{n}= (\alpha_{2},n\alpha_{2})\hspace{50pt}Q_{1}^{n}=
(\alpha_{1},(n-1)\alpha_{1}) \hspace{5pt},\hspace{5pt} Q_{2}^{n}=
(\alpha_{2},n\alpha_{2}) \]\mbox{}\vspace{-20pt}\mbox{}
\[ Q_{3-}^{n}= (\alpha_{3}, -\alpha_{1}+n\alpha_{3})\hspace{50pt}\mbox{\raisebox{5pt}{ or }}\hspace{55pt}
 Q_{3+}^{n}= (\alpha_{3}, +\alpha_{1}+n\alpha_{3}). \]

More recent work \cite{jerome2,piljin1,piljin2} has shown that BPS
states in higher spin multiplets can also exist here. Towers of
these have the non-simple root as magnetic charge, and can be
thought of as bound states of a dyon of each simple root with $n$
units of one root and $n+m$ or $n-m$ $(m>0)$ units of the other,
depending in which weak coupling-limit we are located. Thus for
the highest root the electric charge can range over the whole root
lattice, although only half at any particular point.

\FIGURE[h]{ \centering
\makebox[6cm]{\includegraphics[width=6cm,height=5cm]{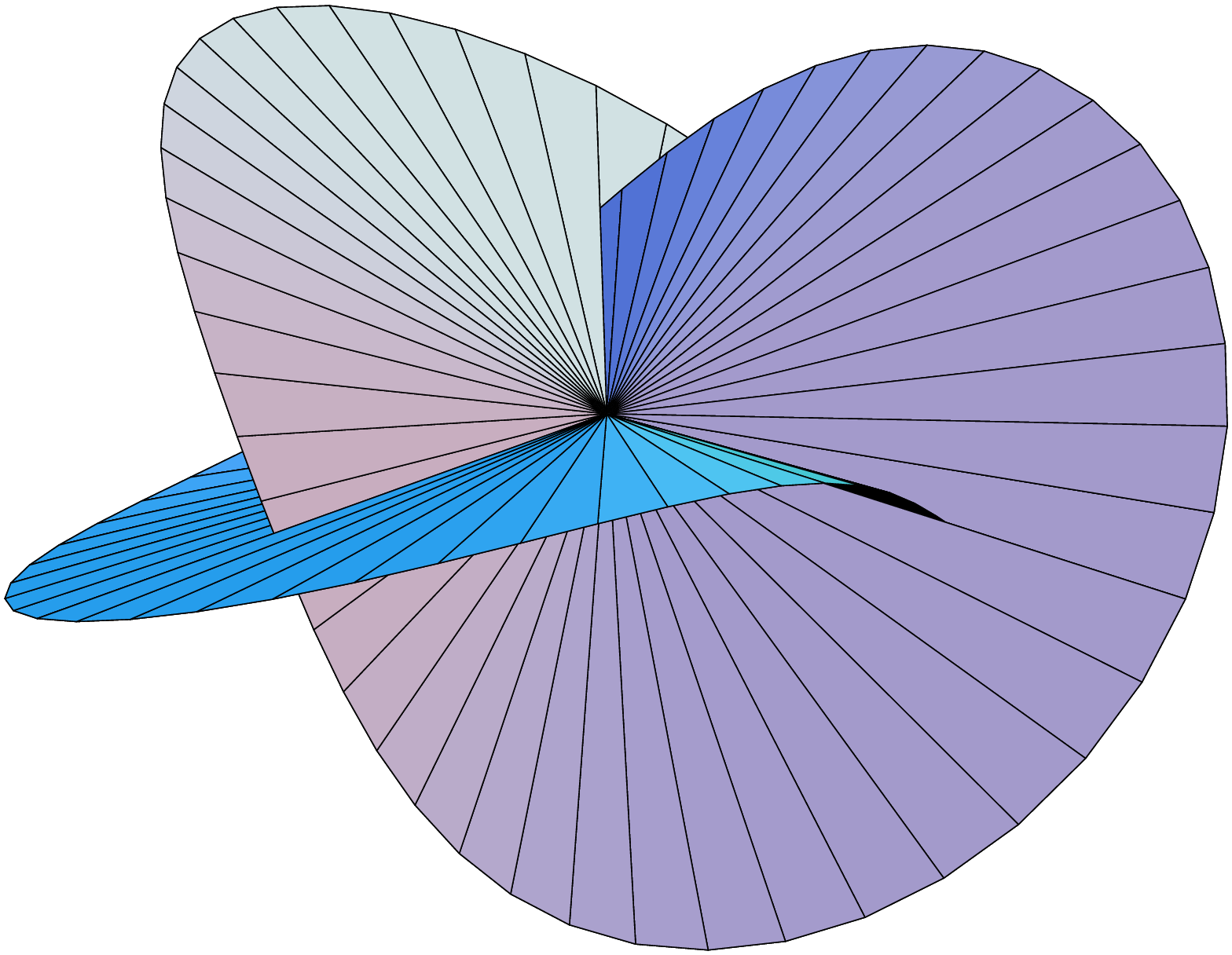}
\put(-115,110){$N_{1}$} \put(-77,95){$N_{3-}$}
\put(-35,110){$N_{2}$} \thicklines \put(-110,70){\vector(0,1){35}}
\put(-72,45){\vector(0,1){45}} \put(-30,70){\vector(0,1){35}}}
\caption{The classical branch cuts form a self-intersecting
surface.} \label{tref1}}

Returning now to the branch cuts, these can be placed between the
trefoil and the origin to form a self-intersecting surface as in
fig.\ \hspace{-3pt}(\ref{tref1}). The transformations, called
monodromies, are labelled $N_{i}$. If we denote by $[(g,q)]$ the
canonical monodromy associated to $(g,q)$, then

\[
N_{i} = [Q_{i}^{n}][Q_{i}^{n-1}] .
\]

Where we use $Q_{3-}$ for $i=3$ (though, in fact we add should
include additional branch cuts as detailed in \cite{me}, such that
we only ever need use simple root monodromies, but we shall ignore
these as we are concentrating solely on when $\alpha_{3}$ is
non-simple).

It is obvious from the diagram that we need a mechanism whereby
$Q_{3+}$'s which are present at the origin, and $Q_{3-}$'s which
are present at infinity, cease to exist in the other region (as
stable BPS bound states). This must happen on the border of when
it is energetically favourable for the bound state to exist, which
in this case is simply when the mass of the bound state is equal
to the sum of the masses of its constituents. This occurs in this
case when our two Higgs fields align. Topologically this
real-codimension one surface, called a curve of marginal stability
(CMS), is a twisted ribbon as in fig.\
\hspace{-3pt}(\ref{ribbon}).

\FIGURE[h]{ \centering
\makebox[6cm]{\includegraphics[width=6cm,height=5cm]{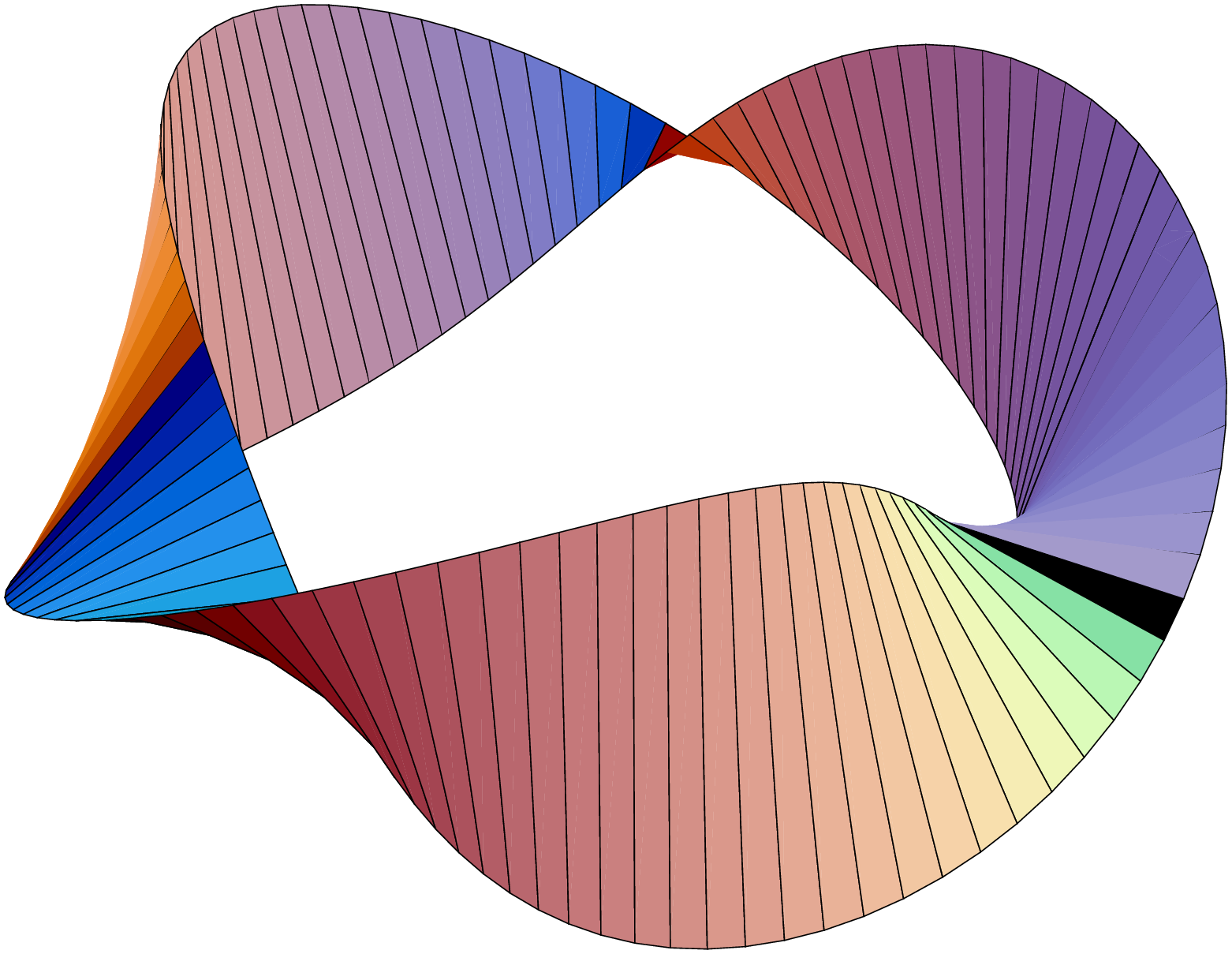}}
\caption{The classical CMS.} \label{ribbon}}

Let us take radial sections of this picture in order to help
visualise things, whilst bearing in mind that we may often need to
rotate this section through $2\pi$ to check consistency (see fig.\
\hspace{-3pt}(\ref{radsec1})).

\FIGURE[ht]{ \centering
\includegraphics[width=10cm]{radsec1.pstex}
\put(-46,96){$N_{1}$} \put(-54,59){$N_{2}$} \caption{We take
radial sections.} \label{radsec1}}

We now modify the model of \cite{me} with the addition of other
such CMS's. Upon these, the states in higher spin multiplets with
electric charge $(n+m)$ or $(n-m)$ units of one simple root decay
to $(n+1)$ or $(n-1)$ together with $(m-1)$ copies of the
appropriate massive gauge boson. These take the shape of just a
small and a large sphere in our model. See fig.\
\hspace{-3pt}(\ref{radsec2}) for the details of how the remaining
dyons are partitioned by these cuts and CMS's into two regions
$3+$ and $3-$.

\FIGURE[ht]{
\makebox[10cm]{\includegraphics[width=8cm]{radsec2.pstex}
\put(-221,122){$N_{1}$} \put(-221,-10){$N_{2}$}
\put(-186,126){$Q_{1}^{n}$} \put(-142,126){$Q_{2}^{n}$}
\put(-97,126){$Q_{3+}^{n}$} \put(-186,163){$Q_{1}^{n-2}$}
\put(-142,163){$Q_{3-}^{n}$} \put(-97,163){$Q_{2}^{n}$}
\put(-186,-8){$Q_{1}^{n}$} \put(-142,-8){$Q_{2}^{n}$}
\put(-97,-8){$Q_{3-}^{n}$} \put(-186,29){$Q_{3+}^{n-2}$}
\put(-142,29){$Q_{2}^{n-2}$} \put(-97,29){$Q_{1}^{n}$}
\put(-110,55){$Q_{3+}^{n}$} \put(-136,101){$Q_{1}^{n}+Q_{2}^{n}$}
\put(2,55){$Q_{1}^{n+2}+Q_{2}^{n}$} \put(2,101){$Q_{3-}^{n}$}
\put(-190,80){\Huge$3+$} \put(35,80){\Huge$3-$}
} \caption{The classical BPS spectra in the $3+$ and $3-$
domains.} \label{radsec2}}

\subsection{Instanton Corrections}

In ${\cal N}=2$ gauge theories non-perturbative corrections are
due to instantons, and are proportional to positive integer powers
of the dynamically generated scale $\Lambda$. At strong coupling
these effects dominate and the model is best expressed in terms of
a dual abelian gauge theory with weak coupling. In the original
theory we are considering 3-spheres of radius $R$. Large
$R/\Lambda$ corresponds to weak coupling. Decreasing from infinite
$R/\Lambda$, the trefoil splits in two, and a new quantum
monodromy is formed between them. This is analogous to the case
for SU$(2)$ where classically one singularity splits into two with
a cut formed between them. Also like in this case, a new, closed
CMS is formed, signalling the loss of stability inside the curve
for a BPS state with a particular charge, in fact for an infinite
number of cases of charge. We call the structure from which the
monodromy $N_{i}$ extends to the origin the $\alpha_{i}$ arm. From
the point of view of the states charged only with $\alpha_{i}$
this is a carbon copy of the situation for SU$(2)$, which is only
to be expected as the dyons are nothing more than SU$(2)$
embeddings along the sub-SU$(2)$ corresponding to this root. This
is illustrated in fig.\ \hspace{-3pt}(\ref{sem10}).

\FIGURE[h]{ \centering
\makebox[12cm]{\includegraphics[width=8cm]{sem10.pstex}
\put(-23,110){$Q_{i}^{n}$, $W_{i}$} \put(-167,109){$Q_{i}^{n}$,
$W_{i}$} \put(-42,66){$Q_{i}^{1}$} \put(-90,66){$Q_{i}^{1}$}
\put(-42,33){$Q_{i}^{2}$} \put(-90,33){$Q_{i}^{0}$}
\put(-185,22){$N_{i}$} \put(-65,96){\scriptsize $Q_{i}^{1}$}
\put(-60,0){\scriptsize $Q_{i}^{2}$} \put(-73,13){\scriptsize
$Q_{i}^{0}$}} \caption{The region where the $\alpha_{i}$ direction
is strongly coupled.} \label{sem10}}

As we rotate our radial section, the two arms rotate around each
other, and the region inside the new curve remains bounded from
the outside. The CMS, which we shall call $L$ for local (and
$L_{i}$ the subset on the $\alpha_{i}$ arm) forms a tube in the
shape of the classical trefoil as in fig.\
\hspace{-3pt}(\ref{lissy}). Inside the tube is the core, a region
of moduli space with a much reduced BPS spectrum. As illustrated
in fig.\ \hspace{-3pt}(\ref{sem10}), on each arm, from a given
direction a single state becomes massless on each trefoil. We have
chosen the branch where $Q_{i}^{1}$ does so at the top. The
spectrum of states with magnetic charge $\alpha_{i}$ in the
hemi-tube along the $\alpha_{i}$ arm will consist only of the two
states which become massless in that section, this is reasonable
as they are light. In \cite{me} we suggested a mechanism whereby
the states in the other two towers are transformed or split into
three (two co-existing at any point) differently charged towers of
states which are relatively light at this arm. This duty is
discharged by the remnant of the ribbon-like CMS. Adding quantum
corrections it splits into an infinite number of curves, which we
call the $G$'s (for global), each delineating the boundary of
stability of one of the $Q_{3\pm}$'s.

\FIGURE[h]{ \centering
\makebox[6cm]{\includegraphics[width=6cm,height=6cm]{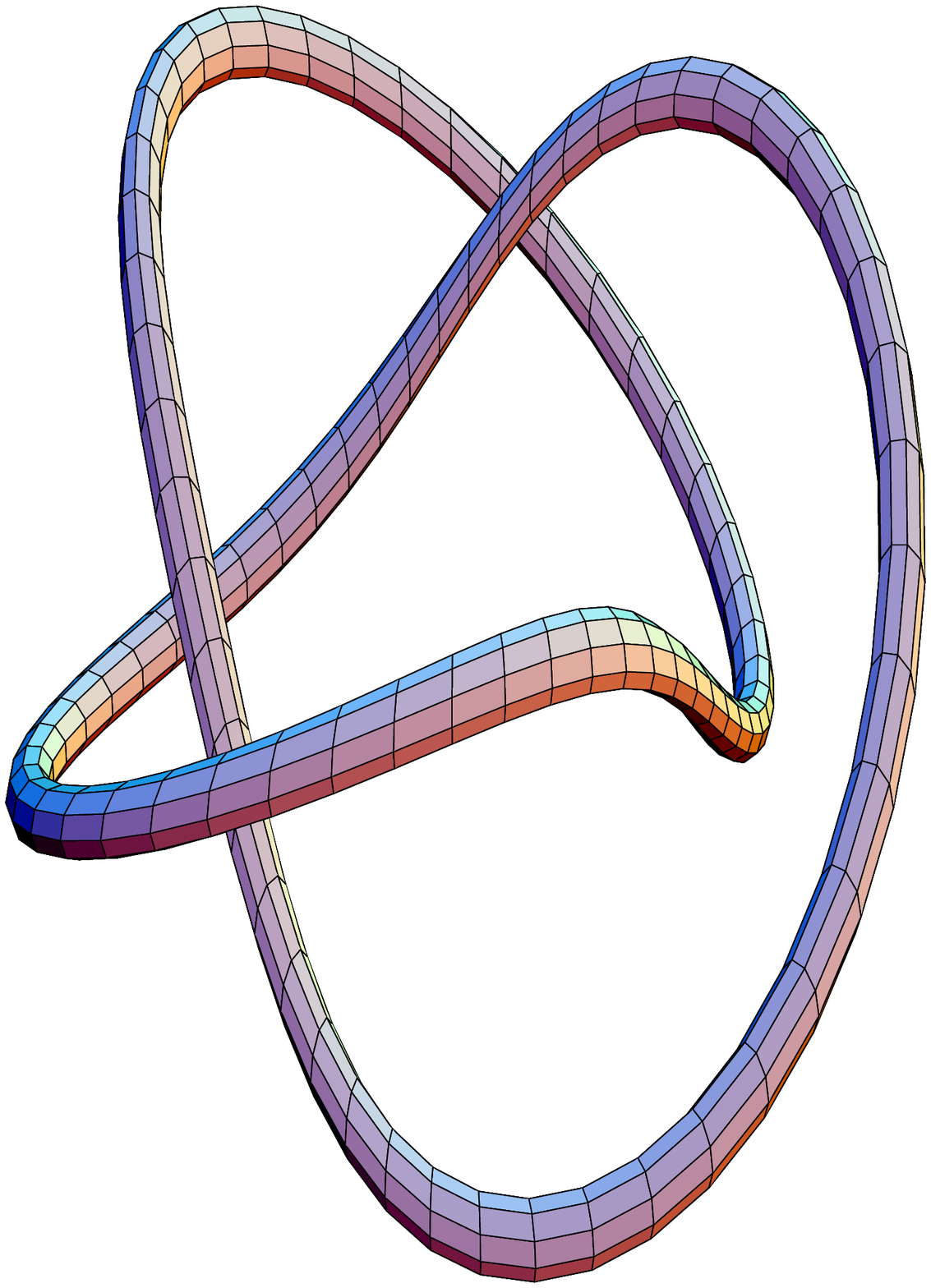}}
\caption{The CMS $L$ which bounds the core.} \label{lissy}}

We define
\begin{eqnarray}
S_{m, i}^{\mbox{-}1}  & = & Q_{i-1}^{2n+1}N_{i}^{m} \n \\ S_{m,
i}^{0} &
= & Q_{i-1}^{2n+2}N_{i}^{m} \n\\
 S_{m, i}^{1} & = &
Q_{i-1}^{2n+3}N_{i}^{m}.\label{sss}
\end{eqnarray}

Within the core on the $\alpha_{i}$ arm we have the states as in
the left hand of fig.\ \hspace{-3pt}(\ref{armicore}). In order to
generalize simply, we would like to change where we put the cuts
$N_{i}$. Instead of always linking to the bottom trefoil, we shall
link to the centre. This now gives us three regions of core with
indices in the new region just one higher than those on the right,
and two greater than those on the left, as depicted in the
right-hand diagram. We could have done this for SU$(2)$ also, and
indeed sometimes it is more natural so to do this, such as when
mapping from the fundamental domain of $\Gamma(2)$.

\FIGURE[h]{ \centering
\makebox[10cm]{\includegraphics[width=10cm]{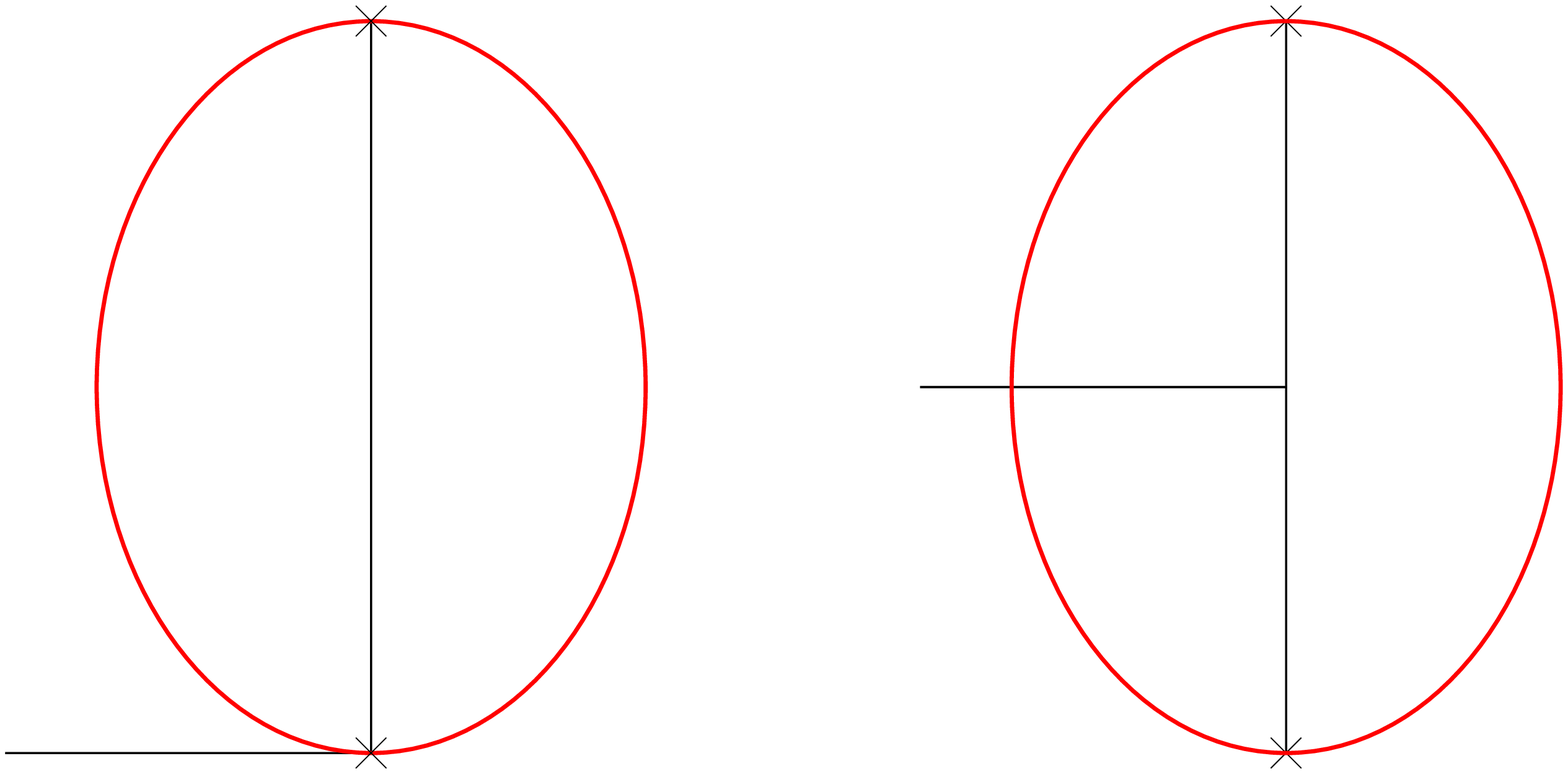}
\put(-245,97){$Q_{i}^{1}$} \put(-245,80){$Q_{i}^{0}$}
\put(-205,97){$Q_{i}^{1}$} \put(-205,80){$Q_{i}^{2}$}
\put(-245,57){$S_{m,i}^{0}$} \put(-205,57){$S_{m,i}^{0}$}
\put(-245,40){$S_{m,i}^{\mbox{-}1}$} \put(-205,40){$S_{m,i}^{1}$}
\put(-37,97){$Q_{i}^{1}$} \put(-37,80){$Q_{i}^{2}$}
\put(-71,117){$Q_{i}^{1}$}
\put(-71,100){$Q_{i}^{0}$}\put(-73,37){$Q_{i}^{2}$}
\put(-73,20){$Q_{i}^{1}$}\put(-96,81){$S_{m,i}^{\mbox{-}1}$}
\put(-74,81){$S_{m,i}^{0}$}\put(-96,54){$S_{m,i}^{1}$}
\put(-74,54){$S_{m,i}^{2}$}\put(-38,54){$S_{m,i}^{0}$}
\put(-38,40){$S_{m,i}^{1}$}}
 \caption{States in the single core
with old and new cuts}\label{armicore}}

If we continue to decrease the radius of our three-sphere,
eventually the trefoils will be so far apart that it will become
impossible for them (and $L$) not to self-intersect. Note that it
is always different arms which intersect, and hence we interpret
this as where both gauge U(1)'s are strongly coupled. We maintain
that the cores cannot just pass through one another with their
intersection supporting the intersection of the two cores for each
arm. Rather, they amalgamate, like a crossroads junction in a
sewer, and concentric spheres of entirely new curves block the
contents of each tube from the other. In the centre of the
junction, inside these new curves, lies a smaller, finite spectrum
of BPS states common to both of the intersecting cores. We call
these spaces the double cores. They exist in the vicinity of both
the $A_{1}$ and $A_{2}$ singularities of the singularity surface.

Although the curves and states listed in \cite{me} are consistent,
we have little way at present of knowing what precisely is
happening at such strong coupling. We now believe that a more
simple scenario is more likely to occur. It involves allowing a
further branch cut $V$ to open up and straddle the intersection of
the tubes. $V$ consists of just the $\alpha_{3}$ Weyl reflection
in both electric and magnetic sectors. We sketch how this may
occur (ignoring the $S$'s) in fig.\ \hspace{-3pt}(\ref{newdcore}).
Note that the curves around the double core now work for an
infinite number of $m$ at the same time, similarly to the $L_{i}$
for the $n$. One might expect a decay into the $S$ of lowest
possible $m$, {\it i.e.}, $S_{2m,i}\rightarrow m
S_{0,i}-(m-1)S_{\mbox{-}2,i},\,\,S_{2m+1,i}\rightarrow m
S_{1,i}-(m-1)S_{\mbox{-}1,i}$. This would leave ten states at each
point of the double core. We believe there should be six. As odd
and even $m$ behave separately, each would still differ in
electric charge by twice $P_{i}=\alpha_{i+1}-\alpha_{i-1}$ ({\it
e.g.\ }\hspace{-3pt} around the $\alpha_{1}$ core we have
$P_{1}=\alpha_{1}+2\alpha_{2}$). In fact they can, and do decay to
states only one $P_{1}$ apart. Four of these are new to us, in the
$\alpha_{1}$ part: $(Q_{2}^{0}+P_{1}),(Q_{2}^{1}+P_{1}),
(\alpha_{3},0)$ and $(\alpha_{3},\alpha_{3})$. The other four in
this area are $S_{\mbox{-}1,1}^{0}, S_{\mbox{-}1,1}^{1},
S_{0,1}^{0}$ and $S_{0,1}^{1}$. The extra four states exist only
inside a spherical `bubble' and lose stability on one curve each.
These mark the boundary of the double core, passing through both
accumulation points, but also extend to touch one of the
singularities of the $\alpha_{i}$ core. In fig.\
\hspace{-3pt}(\ref{dcorev}) we show a double core at one
particular phase of the intersection.

\FIGURE[h]{ \centering
\makebox[13cm]{\includegraphics[width=14cm,height=9cm]{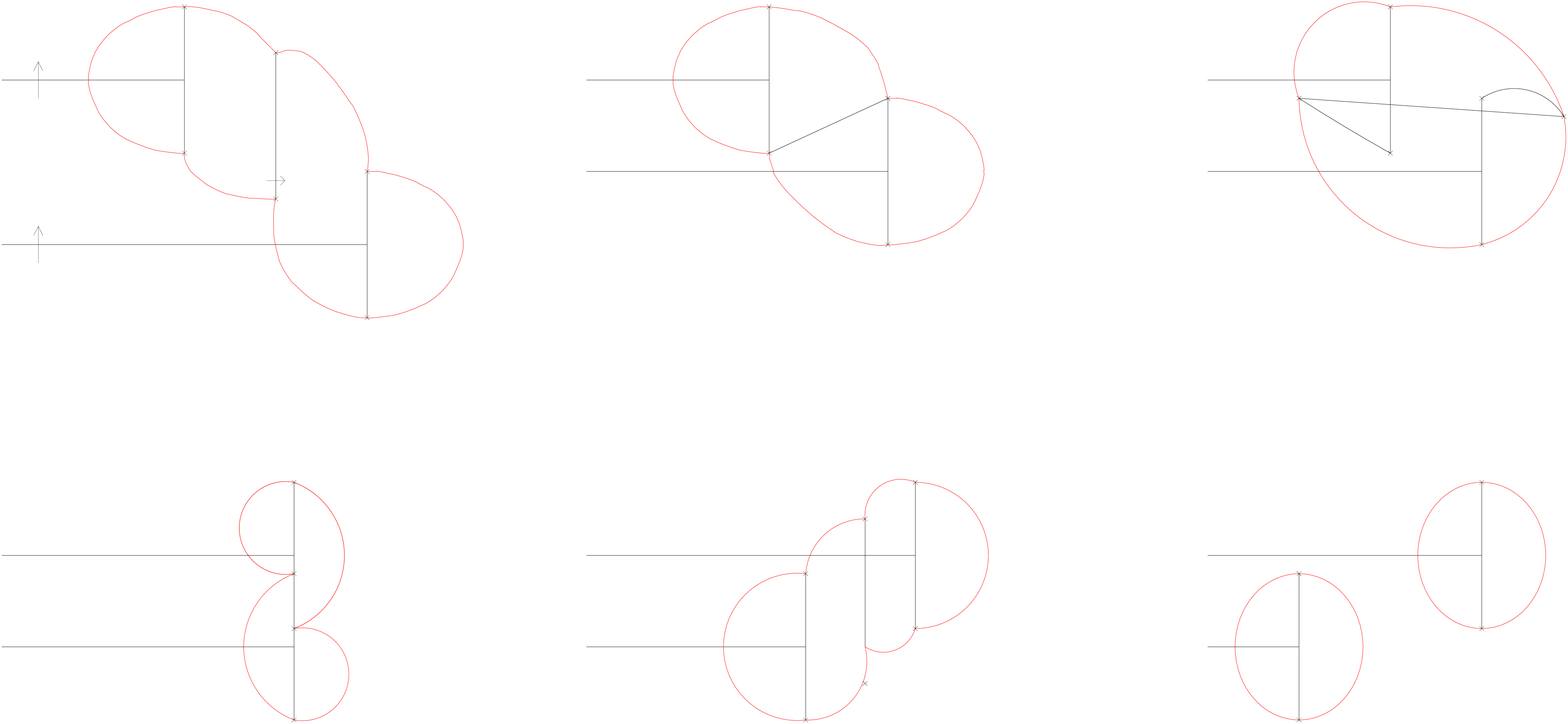}
\put(-394,213){$N_{1}$} \put(-394,155){$N_{2}$}
\put(-337.5,190){$V$}
\put(-364,243){\scriptsize$Q_{1}^{1}$}
\put(-372,233){\scriptsize$Q_{1}^{0}$}
\put(-372,220){\scriptsize$Q_{1}^{2}$}
\put(-364,210){\scriptsize$Q_{1}^{3}$}
\put(-345,226){\scriptsize$Q_{1}^{1}$}
\put(-345,211){\scriptsize$Q_{1}^{2}$}
\put(-321,203){\scriptsize$Q_{2}^{1}$}
\put(-321,188){\scriptsize$Q_{2}^{0}$}
\put(-324,162){\scriptsize$Q_{2}^{3}$}
\put(-317,153){\scriptsize$Q_{2}^{2}$}
\put(-300,172){\scriptsize$Q_{2}^{2}$}
\put(-300,160){\scriptsize$Q_{2}^{1}$}
\put(-216,243){\scriptsize$Q_{1}^{1}$}
\put(-223,233){\scriptsize$Q_{1}^{0}$}
\put(-223,220){\scriptsize$Q_{1}^{2}$}
\put(-216,210){\scriptsize$Q_{1}^{3}$}
\put(-194,235){\scriptsize$Q_{1}^{1}$}
\put(-194,222){\scriptsize$Q_{1}^{2}$}
\put(-197,199.5){\scriptsize$Q_{2}^{0}$}
\put(-186,207){\scriptsize$Q_{2}^{1}$}
\put(-195,186){\scriptsize$Q_{2}^{2}$}
\put(-186,176){\scriptsize$Q_{2}^{3}$}
\put(-168,200){\scriptsize$Q_{2}^{1}$}
\put(-168,188){\scriptsize$Q_{2}^{2}$}
\put(-60,245){\scriptsize$Q_{1}^{1}$}
\put(-67,235){\scriptsize$Q_{1}^{0}$}
\put(-57,222){\scriptsize$Q_{\raisebox{2pt}{\tiny 1}}^{2}$}
\put(-67,222.5){\scriptsize$Q_{\hspace{-2pt}\raisebox{2pt}{\tiny
1}}^{3}$}
\put(-54.5,211){\scriptsize$Q_{\hspace{-1.4pt}3\hspace{-0.5pt}+}^{\hspace{-0.6pt}0}$}
\put(-62,213.5){\scriptsize$\raisebox{1pt}{$Q$}_{\hspace{-2pt}2}^{\raisebox{-2pt}{\tiny\hspace{-0.8pt}1}}$}
\put(-39,241){\scriptsize$Q_{1}^{1}$}
\put(-39,228){\scriptsize$Q_{1}^{2}$}
\put(-63,201){\scriptsize$Q_{2}^{1}$}
\put(-37,201){\scriptsize$Q_{2}^{0}$}
\put(-21,218){\scriptsize$Q_{\hspace{-0.5pt}1}^{\hspace{-0.5pt}2}$}
\put(-12,217.5){\scriptsize$Q_{1}^{3}$}
\put(-17,202){\scriptsize$Q_{2}^{1}$}
\put(-18,188){\scriptsize$Q_{2}^{2}$}
\put(-50,185){\scriptsize$Q_{2}^{1}$}
\put(-37,178){\scriptsize$Q_{2}^{2}$}
\put(-334.5,75){\scriptsize$Q_{1}^{1}$}
\put(-335.5,64){\scriptsize$Q_{1}^{0}$}
\put(-332.5,40){\scriptsize$Q_{\hspace{-1.5pt}2}^{\hspace{-1pt}1}$}
\put(-335,30){\scriptsize$Q_{2}^{0}$}
\put(-335,19.5){\scriptsize$Q_{\hspace{-1.5pt}2}^{\hspace{-0.5pt}3}$}
\put(-333,10){\scriptsize$Q_{\hspace{-1pt}2}^{\hspace{-1pt}2}$}
\put(-323,64){\scriptsize$Q_{1}^{1}$}
\put(-323,52){\scriptsize$Q_{1}^{2}$}
\put(-323,20){\scriptsize$Q_{2}^{1}$}
\put(-323,10){\scriptsize$Q_{2}^{2}$}
\put(-206,43){\scriptsize$Q_{2}^{1}$}
\put(-208,31){\scriptsize$Q_{2}^{0}$}
\put(-208,18){\scriptsize$Q_{2}^{3}$}
\put(-206,7){\scriptsize$Q_{2}^{2}$}
\put(-191,42){\scriptsize$Q_{2}^{1}$}
\put(-191,29){\scriptsize$Q_{2}^{2}$}
\put(-177,48){\scriptsize$Q_{1}^{2}$}
\put(-177,35){\scriptsize$Q_{1}^{3}$}
\put(-177,75){\scriptsize$Q_{1}^{0}$}
\put(-177,64){\scriptsize$Q_{1}^{1}$}
\put(-162,63){\scriptsize$Q_{1}^{1}$}
\put(-162,51){\scriptsize$Q_{1}^{2}$}
\put(-79,41){\scriptsize$Q_{2}^{0}$}
\put(-81,30.5){\scriptsize$Q_{2}^{1}$}
\put(-81,18.5){\scriptsize$Q_{\hspace{-0.5pt}2}^{2}$}
\put(-79,8){\scriptsize$Q_{2}^{3}$}
\put(-66,31){\scriptsize$Q_{2}^{1}$}
\put(-66,19){\scriptsize$Q_{2}^{2}$}
\put(-32.5,73){\scriptsize$Q_{1}^{0}$}
\put(-34.5,62.5){\scriptsize$Q_{1}^{1}$}
\put(-34.5,50.5){\scriptsize$Q_{\hspace{-0.5pt}1}^{2}$}
\put(-32.5,40){\scriptsize$Q_{1}^{3}$}
\put(-19,63){\scriptsize$Q_{1}^{1}$}
\put(-19,51){\scriptsize$Q_{1}^{2}$}
}\caption{How $V$ moves as cores intersect, including the effect
on some states}\label{newdcore}}

\FIGURE[h]{ \centering
\makebox[10cm]{\includegraphics[width=10cm]{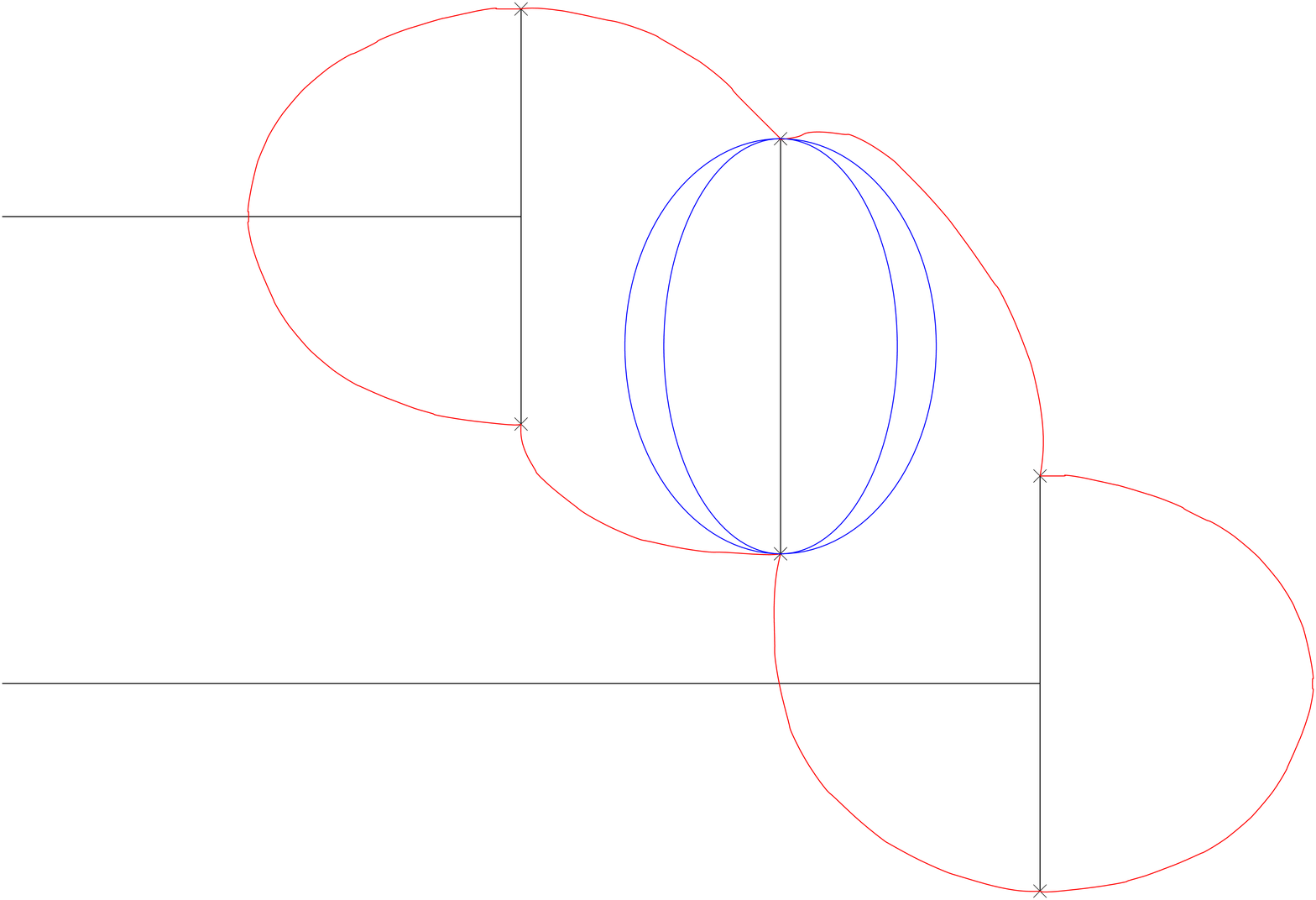}
\put(-195,170){\scriptsize $Q_{1}^{1}$} \put(-210,170){\scriptsize
$Q_{1}^{0}$} \put(-215,155){\scriptsize $S_{m,1}^{0}$}
\put(-196,155){\scriptsize $S_{m,1}^{-1}$}
\put(-195,135){\scriptsize $Q_{1}^{3}$} \put(-210,135){\scriptsize
$Q_{1}^{2}$} \put(-212,120){\scriptsize $S_{m,1}^{1}$}
\put(-193,120){\scriptsize $S_{m,1}^{2}$}
\put(-152,175){\scriptsize $Q_{1}^{2}$} \put(-167,175){\scriptsize
$Q_{1}^{1}$} \put(-165,150){\scriptsize $S_{m,1}^{0}$}
\put(-157,161){\scriptsize $S_{m,1}^{1}$}
\put(-129,151){\scriptsize $Q_{1}^{2}$} \put(-129,139){\scriptsize
$Q_{1}^{1}$} \put(-129,96){\scriptsize $Q_{2}^{0}$}
\put(-129,84){\scriptsize$Q_{2}^{1}$}
\put(-135,125){\scriptsize$Q_{3+}^{1}$}
\put(-135,110){\scriptsize$Q_{3+}^{2}$}
\put(-114,151){\scriptsize $Q_{2}^{1}$} \put(-114,139){\scriptsize
$Q_{2}^{0}$} \put(-114,96){\scriptsize $Q_{1}^{1}$}
\put(-114,84){\scriptsize$Q_{1}^{2}$}
\put(-111,125){\scriptsize$Q_{3-}^{2}$}
\put(-111,110){\scriptsize$Q_{3-}^{3}$}
\put(-89,82){\scriptsize $Q_{2}^{1}$} \put(-75,82){\scriptsize
$Q_{2}^{0}$} \put(-95,67){\scriptsize $S_{m,2}^{0}$}
\put(-100,54){\scriptsize $S_{m,2}^{-1}$}
\put(-99,36){\scriptsize $Q_{2}^{3}$} \put(-84,36){\scriptsize
$Q_{2}^{2}$} \put(-81,10){\scriptsize $S_{m,2}^{2}$}
\put(-94,23){\scriptsize $S_{m,2}^{1}$}
\put(-49,55){\scriptsize $Q_{2}^{1}$} \put(-33,55){\scriptsize
$Q_{2}^{2}$} \put(-51,34){\scriptsize $S_{m,2}^{0}$}
\put(-29,34){\scriptsize $S_{m,2}^{1}$}
}\caption{States in the double core (centre) with monodromy
$V$}\label{dcorev}}

As we continue to shrink our three-sphere smaller than the radius
where the singular points of the singularity curve occur, we find
that the knot has become disentangled into three circles, each
with a core relating to a single arm and all the other
accoutrements of that arm, including the $N_{i}$ monodromy running
in to the centre of this circle. Eventually these circles shrink
to a point then disappear.

\section{One flavour and preliminaries}

\subsection{Matter matters}

The addition of fundamental matter in the SU$(2)$ case is rather
less illustrative of the general scenario than that for pure
vector-multiplets. Seiberg and Witten \cite{sw2} make sure to
include a factor of two in a change of conventions when
hypermultiplets in the fundamental are included. This is actually
the action of the Cartan matrix of SU$(2)$ on the electric sector
of the theory, required because now the electric charges of dyons
and quarks take values in the weight lattice of the group, as
opposed to just the root lattice before. This linear map also
transforms the weights back into a multiple of the roots (as they
are dual) and corresponds to the isogeny of the jacobian of the
auxiliary elliptic curve. It is not necessary for us to change
conventions at all unless we want explicit representations of the
monodromy matrices, and even then not needed if, as we have
previously, used the far-sighted conventions of
\cite{kl1,kl2,klt}.

In SU$(2)$, the weight and root lattice are necessarily collinear,
making it difficult to discern whether a state is a bound state of
a dyon with two quarks with equal and opposite bare mass, or just
the dyon with the next highest charge. Here, SU$(3)$ is more
transparent. SU$(2)$ also has non-standard flavour symmetry. For
$N_{f}$ massless hypermultiplets this symmetry expands from the
usual U$(N_{f})$ to O$(2N_{f})$. Again, SU$(3)$ is free from this
complication.

Following the impressive progress one can make in the pure case of
looking for SU$(2)$ embeddings, charting where they must lead and
the implications, one might be hoping for similar inspiration
here. Sadly, of course, prospective SU$(2)$ weights, lying along
root directions with half-integer spacing, are not SU$(3)$
weights, which lie at the vertices and centroids of the
equilateral triangles formed by the root lattice. Thus we can
infer that searching for embeddings of the states, multiplicities,
monodromies and superconformal points as detailed in
\cite{bf1,bf2,bf3} is folly. Rather, the extra singularities which
become those where very bare-massive quarks become massless stay
well out of the way, for the most part, of the pure SU$(2)$
embeddings and the core.

The new feature of these extra singularities is that upon
traversing a path around them $a$ and $a_{D}$ pick up an integer
multiple of the bare mass of the state which becomes massless at
the singularity. The flavour symmetry in the generic case is a
global U$(1)^{N_{f}}$, and all of the BPS states can be thought of
as carrying additional quantum numbers, having an integer charge
under each of these U$(1)$'s. We can think of these as `merging'
when two bare masses become degenerate, summing the old charges.
We can then extend the canonical labelling of monodromies. A state
with charges $(g, q)_{s}$, with $s=(s_{1}, \dots, s_{N_{f}})$
which becomes massless has the associated monodromy
\[
M_{(g,\, q)_{s}}= \left(
\begin{array}{c|c|c}
1+q\otimes g \mbox{\hspace{5pt}}& \mbox{\hspace{3pt}}q \otimes q \mbox{\hspace{2pt}} & \mbox{\hspace{8pt}}s\otimes q \\[3pt]
\hline \raisebox{13pt}{} \mbox{\hspace{-4pt}}-g \otimes g &
\mbox{\hspace{5pt}}1-g\otimes q \mbox{\hspace{2pt}}&
\mbox{\hspace{2pt}}-s\otimes
g \\[3pt] \hline \raisebox{12pt}{}0 & 0 & \mbox{\hspace{2pt}}1
\end{array}
\right).
\]

Quarks necessarily have one unit of s-charge, but are varied in
their choice of which of the $s_{i}$ to have. Generically, for
each weight of the fundamental representation of the gauge group,
we will find $N_{f}$ hypermultiplets with that electric charge,
each with a different $s_{i}$ being non-zero. One sensible
computational basis of weights is that dual to the simple roots,
which has $(N-1)=2$ elements. It can be more instructive for
SU$(N)$ to use the fact that each of the $N(N-1)$ roots can be
expressed as a different difference of two of the $N$ weights of
the fundamental representation. Physically, this means here that
each gauge boson can be considered a bound state of a quark and
antiquark (of the same flavour).

Each matter hypermultiplet contains a colour triplet of quarks. We
label our three quarks by their electric charge: $\lambda_{1}$,
$\lambda_{2}$ and $\lambda_{3}=-(\lambda_{1}+\lambda_{2})$, where
the lambdas are the three weights of the fundamental,
$\alpha_{1}=\lambda_{2}-\lambda_{3}$ and cyclic permutations. If
flavours have different masses, then we need a further label to
distinguish each set of different mass. We give each an s-charge,
of one, under the U$(1)$ part of the flavour symmetry this subset
possesses. This quantity shows up in the central charge, with $a$
and $a_{D}$, multiplied by the corresponding bare mass (with a
factor of $(1/\sqrt{2})$ which we can safely ignore in this
qualitative approach). This, of course, just has the effect of
adding the bare mass to the central charge, the modulus of the
total now being the mass of a state at a particular point of the
moduli space. Following the example of \cite{bf2} we could label
the s-charges of a state as an $N_{f}$-tuple of subscripts
following the other charges, {\it e.g.},
$(g,q)_{s_{1},\,s_{2},\,\dots,\,s_{N_{f}}}$, but as the charge is
always $+1$ for quarks (and of obvious sign for other states), we
prefer to write, for example, $(g,q)^{ac}$ for
$(g,q)_{10100\dots}$.

We will later require notation to enumerate the quark-dyon bound
states. We will find dyons with magnetic root $\alpha_{1}$ bind to
the weight orthogonal to this, which in our notation is labelled
$\lambda_{2}$. As $\lambda_{2}-\alpha_{1}=\lambda_{3}$, any such
state could also be thought of as a bound state with
$\lambda_{3}$. We shall always label it relating to the first
possibility. We draw a bar over the name of the dyon for each such
quark bound with it, and add the s-charge labels appropriately.
Antiquarks are also apt to form bound states, we then add lower
bars. In a similar way, we add an upper bar for each $\lambda_{1}$
bound with $\alpha_{2}$ dyons, and for each $\lambda_{1}$ with all
$\alpha_{3}$ dyons.

\TABLE{
\begin{tabular}{c|c}
$N_{f}$ & rotation fraction \\
\hline $0$ & $1/3,\, 1/2$ \\
$1$ & $2/5,\,3/5$ \\
$2$ & $1/2,\,3/4$ \\
$3$ & $2/3,\,1$ \\
$4$ & $1,\,1/2$ \\
$5$ & $1,\,1$ \\
\end{tabular}
\caption{duality symmetries}\label{tab0}}

The pure SU$(2)$ moduli space described by quadratic Casimir $u$
has a discrete $\mathbb{Z}_{2}$ duality symmetry, and with $N_{f}$
masses $\mathbb{Z}_{4-N_{f}}$. By this we mean the theories at $u$
and $u e^{2\pi i /(4-N_{f})}$ will be dual descriptions of the
same physical theory. The order of the group is related to the
anomaly-free subgroup of the classical U$(1)_{R}$ symmetry (and
also parity), which is $\mathbb{Z}_{4(4-N_{f})}$ except for zero
flavours where it is $\mathbb{Z}_{2(4-N_{f})}$. Note $u$ has
charge $4$ under this symmetry, thus we have the stated
invariance. For SU$(3)$ we also get such dualities - just replace
$4(4-N_{f})$ with $2(6-N_{f})$, and include $v$ with charge $6$.
Therefore the symmetries occur for simultaneous rotation of $u$
and $v$ by the amounts listed in Table \ref{tab0}. In particular,
singularities of the singular curve, and also the existence of
cores and double cores will respect this symmetry.

\subsection{SU$(2)$ with Generic Masses}

In \cite{bf3}, Bilal and Ferrari describe in much detail and list
a compendium of relevant CMS's for the case of two flavours with
equal real positive bare masses, and a choice of branch cuts
partly decided by the constraints of computational practicality.
They show the consistency of their description and are solidly
backed by the results of their methodical numerical study of
exactly where these CMS's lie. We free ourselves of these
restrictions such that we may present as simple as possible a
picture as we can muster. The case with four flavours is special
and is dealt with in \cite{bf4}.

The moduli space is always a punctured complex plane. We begin
without matter. There exists a core region about the origin. On
the boundary of this are two singularities in which states of
charge $(1,0)$ and $(1,2)$ respectively become massless. Inside
the core, only two states exist at any point, outside, $(1,2n)$
for all $n$, and the vector-multiplet $(0,2)$. From the core to
infinity there exists a branch cut with monodromy
\[ M_{\infty}^{4} =
\left( \begin{array}{cr}
-1 & 1 \\
0 & -1
\end{array} \right)^{4} = \left( \begin{array}{cr}
-1 & 4 \\
0 & -1
\end{array} \right) .\]

We can add up to four hypermultiplets in the fundamental
representation of the gauge group without destroying the
asymptotic freedom of the theory. If we give them bare masses
$m_{i}$, as we take these masses to infinity, the multiplets
decouple and we return to the pure case. Moreover, we expect to
see this changeover to the pure structure occurring first at a
scale small compared to the bare mass, and then increasing to
infinity as the mass does. Thus, for $N_{1}$ masses varying, after
a brief period of bare mass of order $\Lambda$ where the extra
singularity is in the strong coupling region, increasing $|m_{i}|$
it disentangles itself from the $N_{f}$ core structure to leave
behind the $N_{f}-N_{1}$ core structure, and, if we look at the
generic case when all the multiplets have sizeable bare mass, the
pure core structure. We encumber this concept with the pious tag
`purity comes from within'.

\FIGURE[h]{ \centering
\makebox[10cm]{\includegraphics[width=10cm]{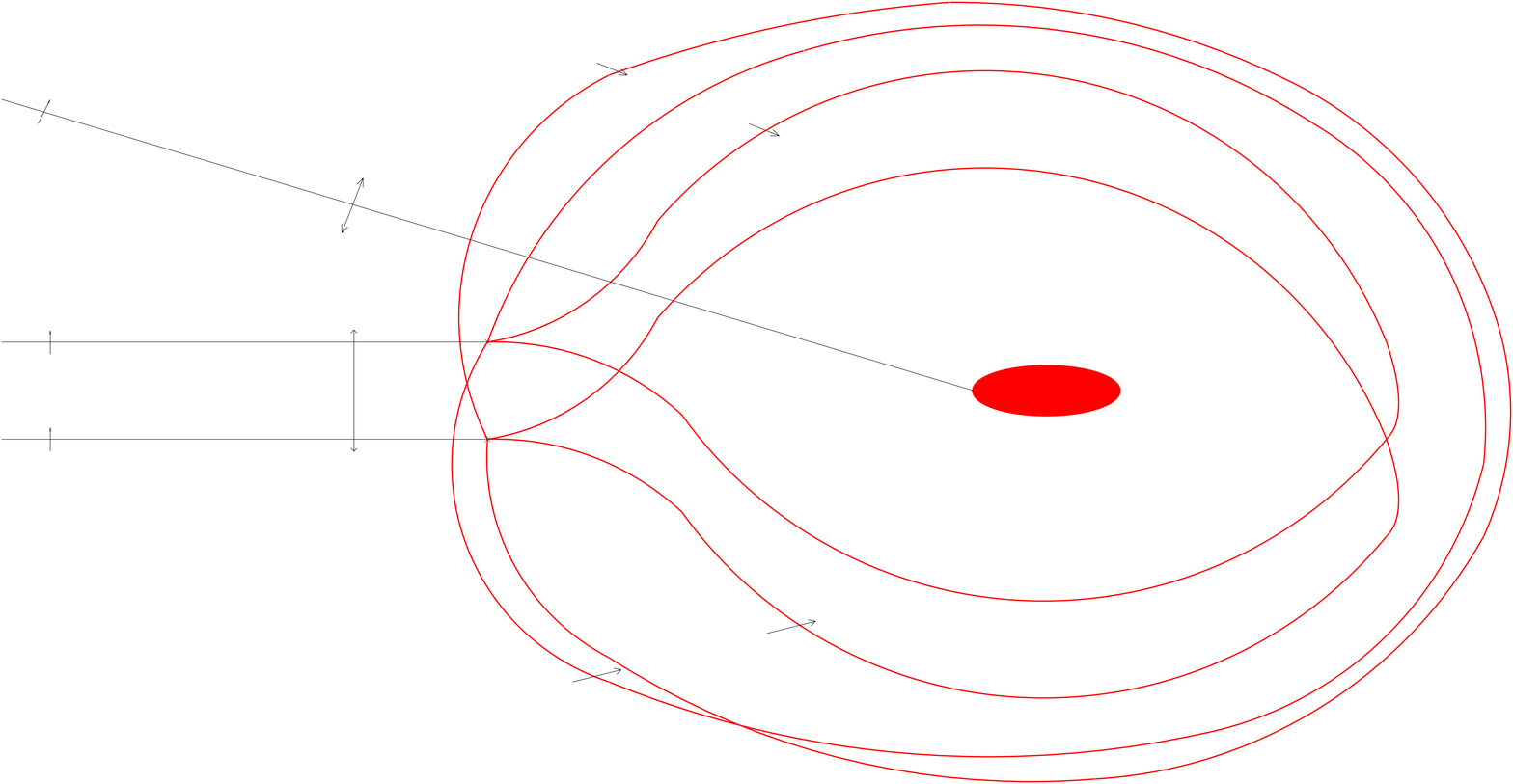}
\put(-289,73){\scriptsize$[(0,1)_{a}]$}
\put(-289,53){\scriptsize$[(0,1)_{b}]$}\put(-287,114){$M_{\infty}^{4}$}
\put(-313,60){$N_{f}\left\{ \begin{array}{c} \\ \\ \\ \end{array}
\right. $}
\put(-263,51){ $Q$} \put(-247,51){$\underline{Q}^{a}$}
\put(-232,51){$\underline{Q}^{b}$} \put(-217,51){$\qdd^{ab}$}
\put(-213,90){ $Q$} \put(-225,90){$\bar{Q}^{a}$}
\put(-240,90){$\bar{Q}^{b}$} \put(-259,90){$\quu^{ab}$}
\put(-260,135){\rotatebox{-18}{$\qdd^{ab}\,\,\underline{Q}^{b}\,\,\underline{Q}^{a}\,\,Q$}}
\put(-158,22){$\underline{Q}^{b}$}
\put(-132,32){\rotatebox{-15}{$Q-(0,1)^{b}$}}
\put(-158,129){$\underline{Q}^{a}$}
\put(-136,115){\rotatebox{10}{$Q-(0,1)^{a}$}}
\put(-186,9){$\qdd^{ab}$}
\put(-178,138){$\qdd^{ab}$}
 }\caption{Generalised sketch of
SU$(2)$ with matter}\label{gensu2}}

If we accept the argument that the set of stable states for
$N_{f}$ flavours is a subset of that for $N_{f}+1$, then
necessarily we find that closed curves of marginal stability must
emanate from each extra singularity, encircling the core, forming
the boundary of the existence domain of states unwanted within.
The monodromies across the branch cuts of each of the $N_{f}$
extra singularities are of the form $[(0,1)_{0,\dots, 1,\dots,0}]$
with the non-trivial s-charge lying in each position once and only
once. These act on a dyon of unit magnetic charge by adding
exactly this charge to the charges of the dyon. In this way we see
the existence of quark-dyon bound states which carry s-charge.
Note that there will exist one of the closed CMS's upon which this
bound state is no longer stable (so one cannot repeatedly encircle
one of these singularities and get a bound state of arbitrarily
high s-charge). Quarks are not affected by these monodromies, so
they commute with each other. Combined with $M_{\infty}^{4}$ we
get a total monodromy that would be $M_{\infty}^{4-N_{f}}$, but it
includes one each of the s-charges. Notice that $M_{\infty}^{4}$
includes the transformation that Seiberg and Witten \cite{sw2}
call $P$ which negates $a$ and $a_{D}$, and hence in our notation
negates all the s-charges. Rotating a dyon about infinity, we
would first pick up one each of positive s-charges, then cross
$M_{\infty}^{4}$ to see them become all $-1$'s. Recrossing the
quark monodromies we would gain charge again to get back to zero.
Thus we do not get arbitrarily high s-charge from this rotation
either. From a glance at fig.\ \hspace{-3pt}(\ref{gensu2}) we see
that at any point only $2^{N_{f}}$ different s-charge combinations
are allowed. In one of the two largest spaces it is the case that
every s-charge can (independently) be $0$ or $+1$; in the other,
that it can be $0$ or $-1$. Between the quark monodromies we get
mixtures.

\subsection{SU$(3)$ with one flavour}

We know that for higher gauge groups the negation property of
$M_{\infty}^{4}$ generalises as a Weyl reflection. In SU$(3)$ each
of the $N_{i}$ makes such a transformation, effectively negating
$a$ and $a_{D}$ in the $\alpha_{i}$ direction, and swapping
without negation, the other two root directions. We need the extra
monodromy we get when we add one flavour to interact with these,
as we saw for SU$(2)$, to restrict the s-charges to $\{ 0,\pm 1 \}
$. The position of the extra singularity for zero bare mass is
easily calculated and is found to be a circle inside the torus on
which the classical trefoil lies. We take the branch cut from this
circle to extend to its centre at the origin. We should then take
the monodromy to be $[(0,\lambda_{3})_{1}]$ as this is weight of
the fundamental representation which is orthogonal to
$\alpha_{3}$, (the 3's only coincidentally corresponding --- they
are the only things that are not $1$ or $2$) the only root in the
fundamental Weyl chamber, to which the dominant principal
direction of the Higgs fields is restricted for this discussion.
As can be seen from fig.\ \hspace{-3pt}(\ref{lissm}), we now have
three regions and an expanded BPS spectrum within each. $Q_{3}$'s
exist only barless or with upper bars. The $\bar{Q}_{2}$'s exist
in the $3-$ and the upper of the two $3+$ regions, and may move
freely between them. Instead, however, $\underline{Q}_{2}$'s exist
in the lower $3+$, and the both of them are marginally stable to
$Q_{2}$'s and $\pm(0,\lambda_{1})_{1}$  (this couldn't happen
simultaneously in the quantum case, but we have extreme degeneracy
in this limit). Similarly $\underbar{Q}_{1}$'s exist only in the
upper $3+$.

\FIGURE[h]{ \centering
\makebox[10cm]{\includegraphics[width=10cm]{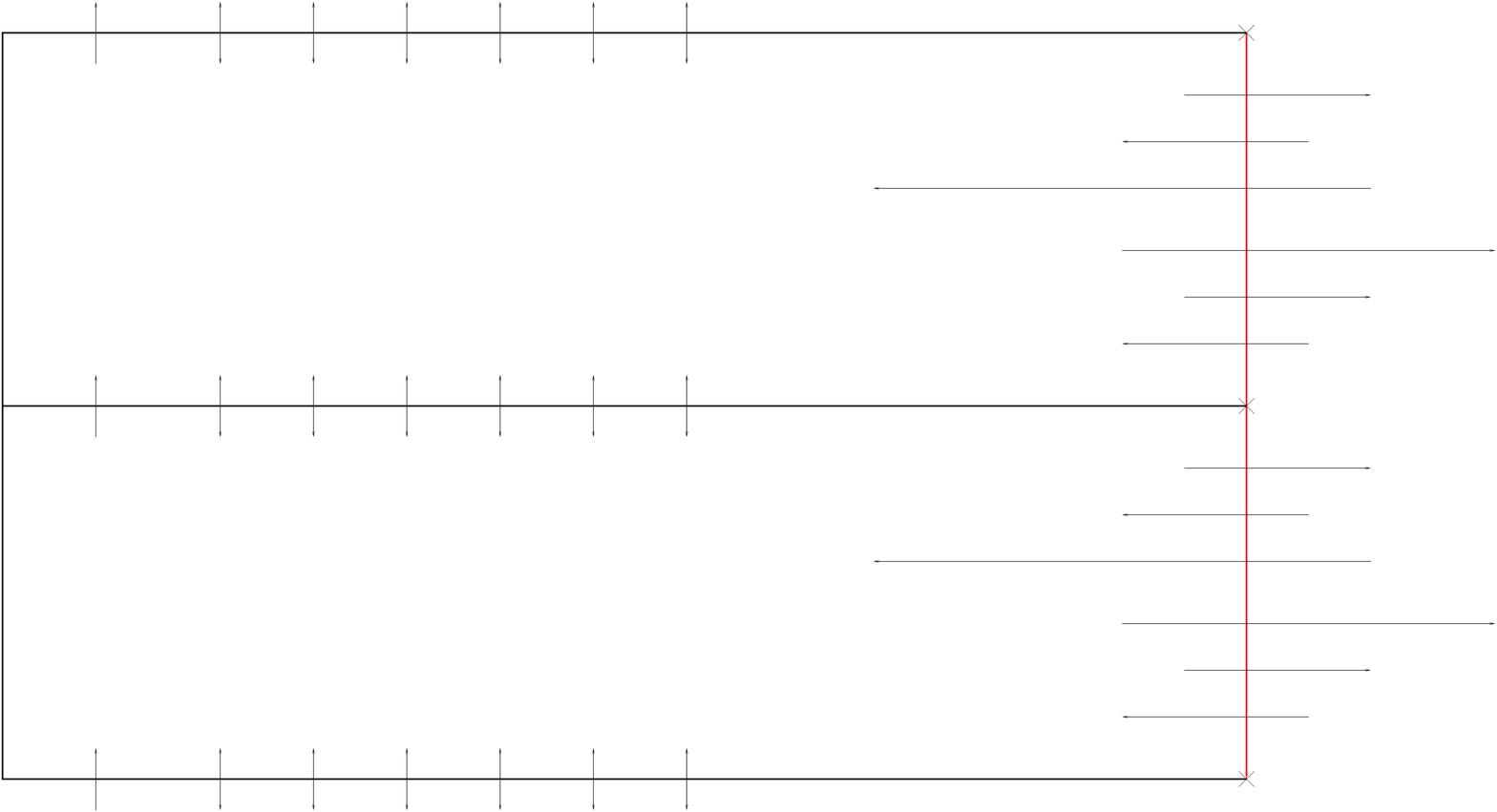}
\put(-273,132){$N_{1}$}
\put(-275,62){$[\lambda_{3}]$}\put(-273,-9){$N_{2}$}
\put(-245,156){\scriptsize$Q_{1}^{n}$}
\put(-227.7,156){\scriptsize$Q_{2}^{n}$}
\put(-210,156){\scriptsize$Q_{3-}^{n}$}
\put(-192,156){\scriptsize$\bar{Q}_{1}^{n}$}
\put(-174.5,156){\scriptsize$\bar{Q}_{2}^{n}$}
\put(-157,156){\scriptsize$\bar{Q}_{3-}^{n}$}
\put(-246,134){\scriptsize$Q_{1}^{\raisebox{3pt}{\tiny$\hspace{-2pt}n\hspace{-2.2pt}+\hspace{-2.2pt}2$}}$}
\put(-228.7,134){\scriptsize$Q_{3+}^{n}$}
\put(-211,134){\scriptsize$Q_{2}^{n}$}
\put(-193,134){\scriptsize$\underline{Q}_{1}^{\raisebox{3pt}{\tiny$\hspace{-2pt}n\hspace{-2.2pt}+\hspace{-2.2pt}3$}}$}
\put(-175.5,134){\scriptsize$\bar{Q}_{3-}^{n}$}
\put(-158,134){\scriptsize$\bar{Q}_{2}^{n}$}
\put(-246,-8){\scriptsize$Q_{1}^{n}$}
\put(-228.7,-8){\scriptsize$Q_{2}^{n}$}
\put(-211,-8){\scriptsize$Q_{3-}^{n}$}
\put(-193,-8){\scriptsize$\bar{Q}_{1}^{n}$}
\put(-175.5,-8){\scriptsize$\bar{Q}_{2}^{n}$}
\put(-158,-8){\scriptsize$\bar{Q}_{3-}^{n}$}
\put(-245,16){\scriptsize$Q_{3+}^{\raisebox{3pt}{\tiny$\hspace{-2pt}n\hspace{-2.2pt}-\hspace{-2.2pt}2$}}$}
\put(-227.7,16){\scriptsize$Q_{2}^{\raisebox{3pt}{\tiny$\hspace{-2pt}n\hspace{-2.2pt}-\hspace{-2.2pt}2$}}$}
\put(-210,16){\scriptsize$Q_{1}^{\raisebox{3pt}{\tiny$\hspace{-1pt}n$}}$}
\put(-192,16){\scriptsize$\bar{Q}_{3+}^{\raisebox{3pt}{\tiny$\hspace{-1pt}n\hspace{-2.2pt}-\hspace{-2.2pt}1$}}$}
\put(-174.5,16){\scriptsize$\underline{Q}_{2}^{\raisebox{3pt}{\tiny$\hspace{-1pt}n\hspace{-2.2pt}-\hspace{-2.2pt}3$}}$}
\put(-157,16){\scriptsize$\bar{Q}_{1}^{\raisebox{3pt}{\tiny$\hspace{-1pt}n\hspace{-2.2pt}-\hspace{-2.2pt}1$}}$}
\put(-245,87){\scriptsize$Q_{1}^{n}$}
\put(-227.7,87){\scriptsize$Q_{2}^{n}$}
\put(-210,87){\scriptsize$Q_{3+}^{n}$}
\put(-192,87){\scriptsize$\underline{Q}_{1}^{n}$}
\put(-174.5,87){\scriptsize$\bar{Q}_{2}^{n}$}
\put(-157,87){\scriptsize$\bar{Q}_{3+}^{n}$}
\put(-246,63){\scriptsize$\bar{Q}_{1}^{\raisebox{3pt}{\tiny$\hspace{-1pt}n\hspace{-2.2pt}-\hspace{-2.2pt}1$}}$}
\put(-228.7,63){\scriptsize$\underline{Q}_{2}^{\raisebox{3pt}{\tiny$\hspace{-2pt}n\hspace{-2.2pt}-\hspace{-2.2pt}1$}}$}
\put(-211,63){\scriptsize$Q_{3+}^{\raisebox{3pt}{\tiny$\hspace{-1pt}n$}}$}
\put(-193,63){\scriptsize$Q_{1}^{\raisebox{3pt}{\tiny$\hspace{-2pt}n\hspace{-2.2pt}-\hspace{-2.2pt}1$}}$}
\put(-175.5,63){\scriptsize$Q_{2}^{\raisebox{3pt}{\tiny$\hspace{-1pt}n\hspace{-2.2pt}-\hspace{-2.2pt}1$}}$}
\put(-158,63){\scriptsize$\bar{Q}_{3+}^{\raisebox{3pt}{\tiny$\hspace{-1pt}n$}}$}
\put(-20,134){\scriptsize$Q_{1}^{n+2}\hspace{-4pt}+\hspace{-2pt}Q_{2}^{n}$}
\put(-75,134){\scriptsize$Q_{3+}^{n}$}
\put(-106,125){\scriptsize$Q_{1}^{n}+Q_{2}^{n}$}
\put(-33,125){\scriptsize$Q_{3-}^{n}$}
\put(-20,116){\scriptsize$\bar{Q}_{1}^{n}$}
\put(-150,116){\scriptsize$Q_{1}^{n}+\lambda_{2}$}
\put(2,105){\scriptsize$Q_{1}^{n}\hspace{-3pt}-\hspace{-3pt}\lambda_{2}$}
\put(-82,105){\scriptsize$\underline{Q}_{1}^{n}$}
\put(-20,95){\scriptsize$Q_{1}^{n+2}\hspace{-4pt}+\hspace{-2pt}\bar{Q}_{2}^{n}$}
\put(-74,95){\scriptsize$\bar{Q}_{3+}^{n}$}
\put(-106,86){\scriptsize$Q_{1}^{n}+\bar{Q}_{2}^{n}$}
\put(-33,86){\scriptsize$\bar{Q}_{3-}^{n}$}
\put(-20,64){\scriptsize$Q_{1}^{n+2}\hspace{-4pt}+\hspace{-2pt}Q_{2}^{n}$}
\put(-75,64){\scriptsize$Q_{3+}^{n}$}
\put(-106,55){\scriptsize$Q_{1}^{n}+Q_{2}^{n}$}
\put(-33,55){\scriptsize$Q_{3-}^{n}$}
\put(-20,46){\scriptsize$\bar{Q}_{2}^{n}$}
\put(-150,46){\scriptsize$Q_{2}^{n}+\lambda_{1}$}
\put(2,35){\scriptsize$Q_{2}^{n}\hspace{-3pt}-\hspace{-3pt}\lambda_{1}$}
\put(-82,35){\scriptsize$\underline{Q}_{2}^{n}$}
\put(-20,25){\scriptsize$\bar{Q}_{1}^{n+1}\hspace{-4pt}+\hspace{-2pt}Q_{2}^{n-1}$}
\put(-74,25){\scriptsize$\bar{Q}_{3+}^{n}$}
\put(-119,16){\scriptsize$\bar{Q}_{1}^{n-1}\hspace{-4pt}+\hspace{-2pt}Q_{2}^{n-1}$}
\put(-33,16){\scriptsize$\bar{Q}_{3-}^{n}$}
} \caption{The picture for one flavour at
$\Lambda/R=0$}\label{lissm}}

\subsection{Non-perturbative Effects: Small $\Lambda$}

We now add in non-perturbative effects, staying in the regime
$R\gg \Lambda$ where these are small.
 The extra monodromy changes many things. The global curves which
in the pure case partitioned the space into two now pass through
this extra branch cut and are altered accordingly. They cannot now
partition all the varied states we have, and one further set of
curves is necessary to complete the picture, but we shall see that
they do so for all the $Q_{3}$'s. Without the matter, in \cite{me}
we showed that in our usual radial section just two small segments
were needed to generate all the curves. Now we need several more.
The picture which works is illustrated in fig.\
\hspace{-3pt}(\ref{gnf1}), where both left acting and right acting
curves that we had included before are present, bending to the
outside of the singularity. A global clockwise rotation of the
arms by three half twists, also including the local half-twist in
one core, returns our radial section to the same place but leaves
the former curves in a different position, now passing to the
inside of the singularity, through the monodromy. Repeating, they
return outside, but with different index $n\rightarrow n-5$. Thus
five copies of the two original curves are needed to generate all
$n$. This is all that is necessary to partition all the $Q_{3}$
and $\bar{Q}_{3}$'s. At the cores we now have a whole set of $G$
as in the pure case, and also a whole set of $\bar{G}$'s which act
on the $\bar{Q}_{i}$'s in the same way (all the states are then
upper-barred except the ones of the same type as the arm around
which we are twisting). These limit what gets into the core to two
copies of what was there in the pure case.

\FIGURE[h]{ \centering
\makebox[10cm]{\includegraphics[width=8cm]{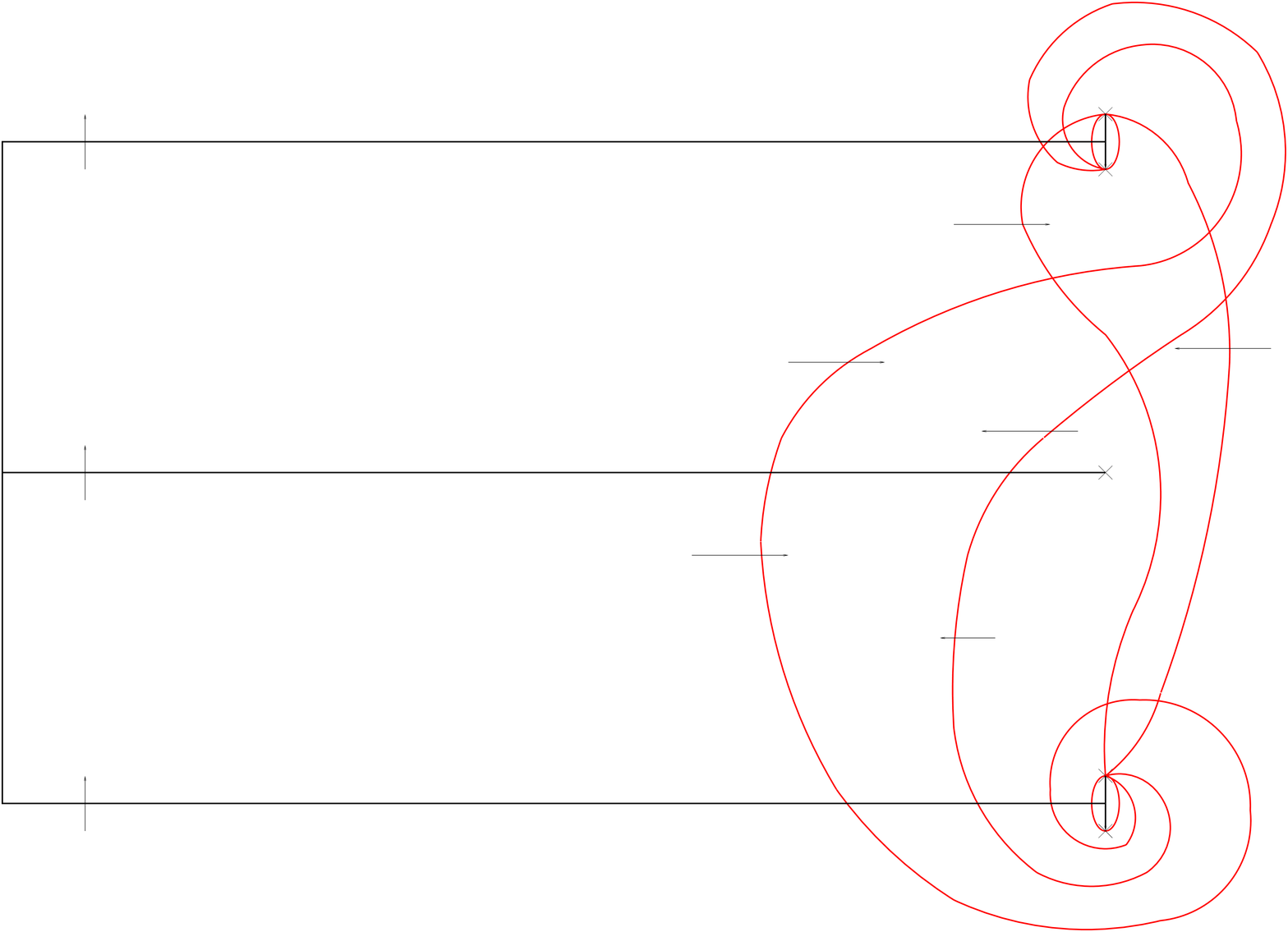}
\put(-218,8){$N_{2}$} \put(-220,66){$[\lambda_{3}]$}
\put(-218,126){$N_{1}$}
\put(-73,123){\scriptsize$Q_{3+}^{1}$}
\put(-41,123){\scriptsize$Q_{1}^{3}Q_{2}^{1}$}
\put(-104,98){\scriptsize$\bar{Q}_{3+}^{\mbox{-}2}$}
\put(-70,98){\scriptsize$Q_{1}^{0}\bar{Q}_{2}^{\mbox{-}2}$}
\put(-119,64){\scriptsize$\bar{Q}_{3+}^{\mbox{-}2}$}
\put(-87,64){\scriptsize$\bar{Q}_{1}^{\mbox{-}1}Q_{2}^{\mbox{-}3}$}
\put(-50,50){\scriptsize$\bar{Q}_{3\mbox{-}}^{0}$}
\put(-87,50){\scriptsize$\bar{Q}_{1}^{\mbox{-}1}Q_{2}^{\mbox{-}1}$}
\put(-36.5,85){\scriptsize$\bar{Q}_{3\mbox{-}}^{0}$}
\put(-77,85){\scriptsize$\bar{Q}_{2}^{0}Q_{1}^{0}$}
\put(-2,101){\scriptsize$Q_{3\mbox{-}}^{1}$}
\put(-42,101){\scriptsize$Q_{2}^{1}Q_{1}^{1}$}
}\caption{Generators of representatives of the $G$'s and
$\bar{G}$'s, which partition the $Q_{3}$ and
$Q_{3}$'s}\label{gnf1}}

\FIGURE[h]{ \centering
\makebox[13cm]{\includegraphics[width=13cm]{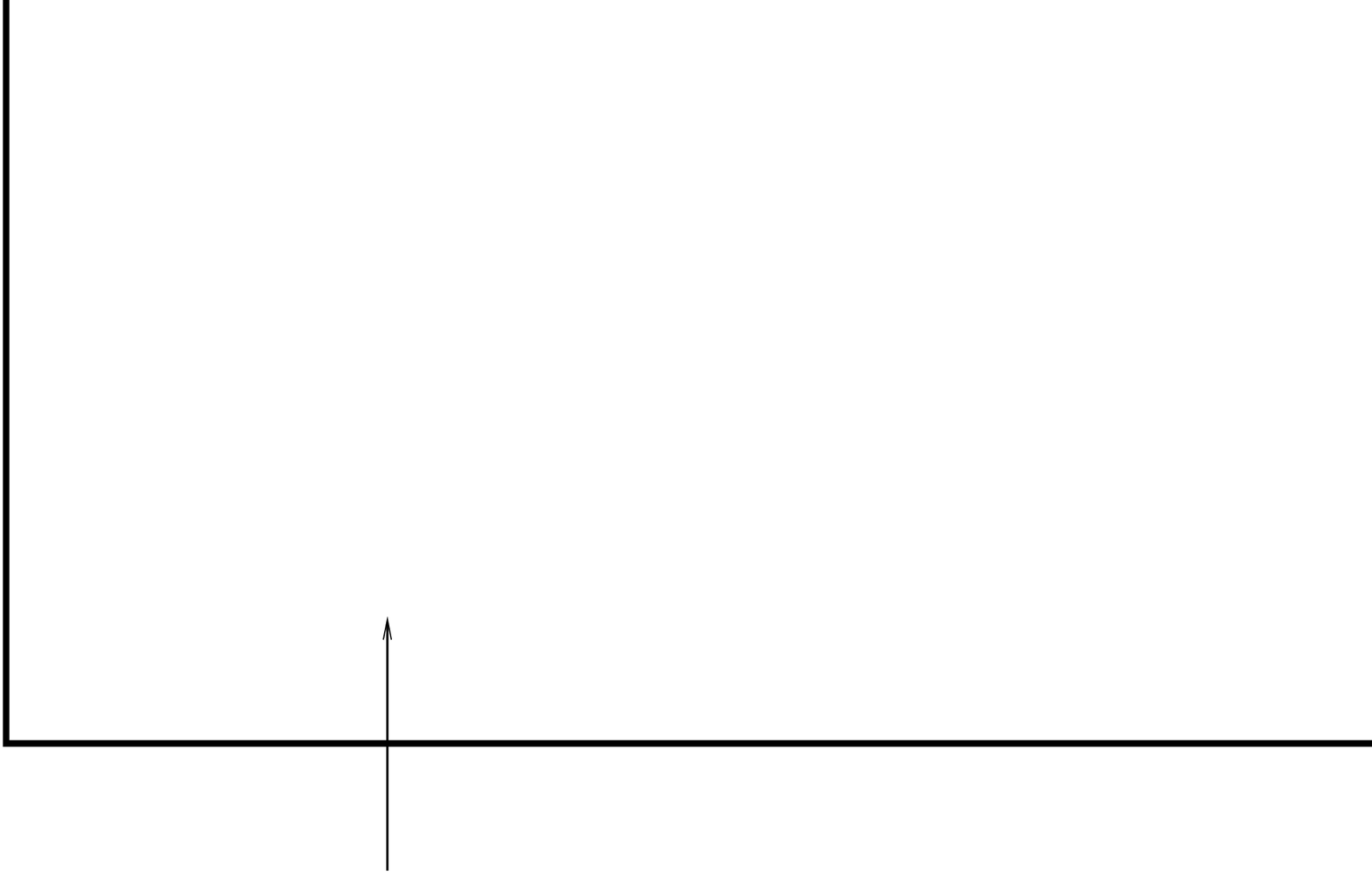}
\put(-335,80){\scriptsize$Q_{2}$}
\put(-316,80){\scriptsize$\lambda_{1}\bar{Q}_{2}$}
\put(-321,101){\scriptsize$\bar{Q}_{3+}$}
\put(-303,94){\scriptsize$Q_{3+}$}
\put(-300,101){\scriptsize$\lambda_{2}$}
\put(-335,70){\scriptsize$\underline{Q}_{2}$}
\put(-316,70){\scriptsize$\lambda_{1}Q_{2}$}
\put(-323,57){\scriptsize$\bar{Q}_{3+}$}
\put(-305,58){\scriptsize$Q_{3+}$}
\put(-300.5,50.5){\scriptsize$\lambda_{2}$}
\put(-290.5,83){\scriptsize$\underline{Q}_{1}$}
\put(-282,93){\scriptsize$Q_{1}$}
\put(-276.5,85){\scriptsize$\lambda_{3}$}
\put(-289.5,68.5){\scriptsize$Q_{1}$}
\put(-281,55){\scriptsize$\bar{Q}_{1}$}
\put(-276.5,62){\scriptsize$\lambda_{3}$}
\put(-249.5,92){\scriptsize$\bar{Q}_{2}$}
\put(-272,103){\scriptsize$Q_{2}$}
\put(-266.5,95){\scriptsize$\lambda_{3}$}
\put(-248.5,59){\scriptsize$\bar{Q}_{2}$}
\put(-262,47){\scriptsize$Q_{2}$}
\put(-266,55){\scriptsize$\lambda_{3}$}
\put(-227,71){\scriptsize$Q_{3\mbox{-}}$}
\put(-258,71){\scriptsize$\lambda_{1}\bar{Q}_{3\mbox{-}}$}
\put(-213,84){\scriptsize$\bar{Q}_{1}$}
\put(-240,84){\scriptsize$\lambda_{2}Q_{1}$}
\put(-119,81){\scriptsize$\underline{Q}_{1}$}
\put(-101.5,81){\scriptsize$\lambda_{2}Q_{1}$}
\put(-107,100){\scriptsize$\bar{Q}_{\hspace{-1pt}3+}$}
\put(-88,97){\scriptsize$Q_{3+}$}
\put(-87,105){\scriptsize$\lambda_{1}$}
\put(-119,69){\scriptsize$Q_{1}$}
\put(-101,69){\scriptsize$\lambda_{2}\bar{Q}_{1}$}
\put(-105.5,56.5){\scriptsize$\bar{Q}_{3+}$}
\put(-88.5,58){\scriptsize$Q_{3+}$}
\put(-84.5,50.5){\scriptsize$\lambda_{1}$}
\put(-77,83){\scriptsize$Q_{2}$}
\put(-64,94){\scriptsize$\bar{Q}_{2}$}
\put(-58.5,86){\scriptsize$\lambda_{3}$}
\put(-76.5,70){\scriptsize$\underline{Q}_{2}$}
\put(-64,60){\scriptsize$Q_{2}$}
\put(-59,67){\scriptsize$\lambda_{3}$}
\put(-35,91.5){\scriptsize$\bar{Q}_{1}$}
\put(-49,103){\scriptsize$Q_{1}$}
\put(-51.5,95.5){\scriptsize$\lambda_{3}$}
\put(-35.5,56){\scriptsize$\bar{Q}_{1}$}
\put(-49,47){\scriptsize$Q_{1}$}
\put(-52.5,55){\scriptsize$\lambda_{3}$}
\put(-13.5,71){\scriptsize$Q_{3\mbox{-}}$}
\put(-42,71){\scriptsize$\lambda_{1}\bar{Q}_{3\mbox{-}}$}
\put(0,84){\scriptsize$\bar{Q}_{2}$}
\put(-26.5,84){\scriptsize$Q_{2}\lambda_{2}$}
\put(-364,111){$N_{1}$} \put(-367,64){$[\lambda_{3}]$}
\put(-364,18){$N_{2}$}
\put(-148,111){$N_{1}$} \put(-151,64){$[\lambda_{3}]$}
\put(-148,18){$N_{2}$}
}\caption{Sketch of the new curves needed to partition the
extra states}\label{sweepnf1a}}

There remains, of course, the curves which partition upper- and
lower-barred $Q_{1}$ and $Q_{2}$'s, a simplified version of which
we show pictorially in fig.\ \hspace{-3pt}(\ref{sweepnf1a}). The
regions around the two cores are devoid of lower-barred states,
having a smaller spectrum than the weak coupling one. Remember
that the $Q_{3}$'s and $\bar{Q}_{3}$'s in these regions are
partitioned by $G$ and $\bar{G}$'s. This time we also need five
copies in order to range over all $n$. All three quarks must be
present throughout this region of the moduli space, including
inside the cores. The fact that they should not pass through a
core (other than their own) return and pass through again in the
same direction alerts us to the fact that the situation is more
complicated than we have drawn. Naturally curves involving, say,
$Q_{1}$'s as constituent states, which arc around the $\alpha_{1}$
core will actually touch the core after wrapping around the
requisite number of times (then crossing back through $N_{1}$ the
same amount of times to return to the same global picture as fig.\
\hspace{-3pt}(\ref{sweepnf1a})). Thus a more accurate picture
would be fig.\ \hspace{-3pt}(\ref{sweepnf1b}), where we have drawn
the intersections for a particular choice of $n$. Looking at the
core in detail in fig.\ \hspace{-3pt}(\ref{corenf1s}), where we
have included just the states which transform to $\lambda_{2}$ and
$\lambda_{3}$ through the quantum cut, we see how they cannot
leave the core and come back in the other side. This allows for
the existence of all the quarks in the core to be consistent. The
presence of these further states is no problem.

\FIGURE[h]{ \centering
\makebox[13cm]{\includegraphics[width=13cm]{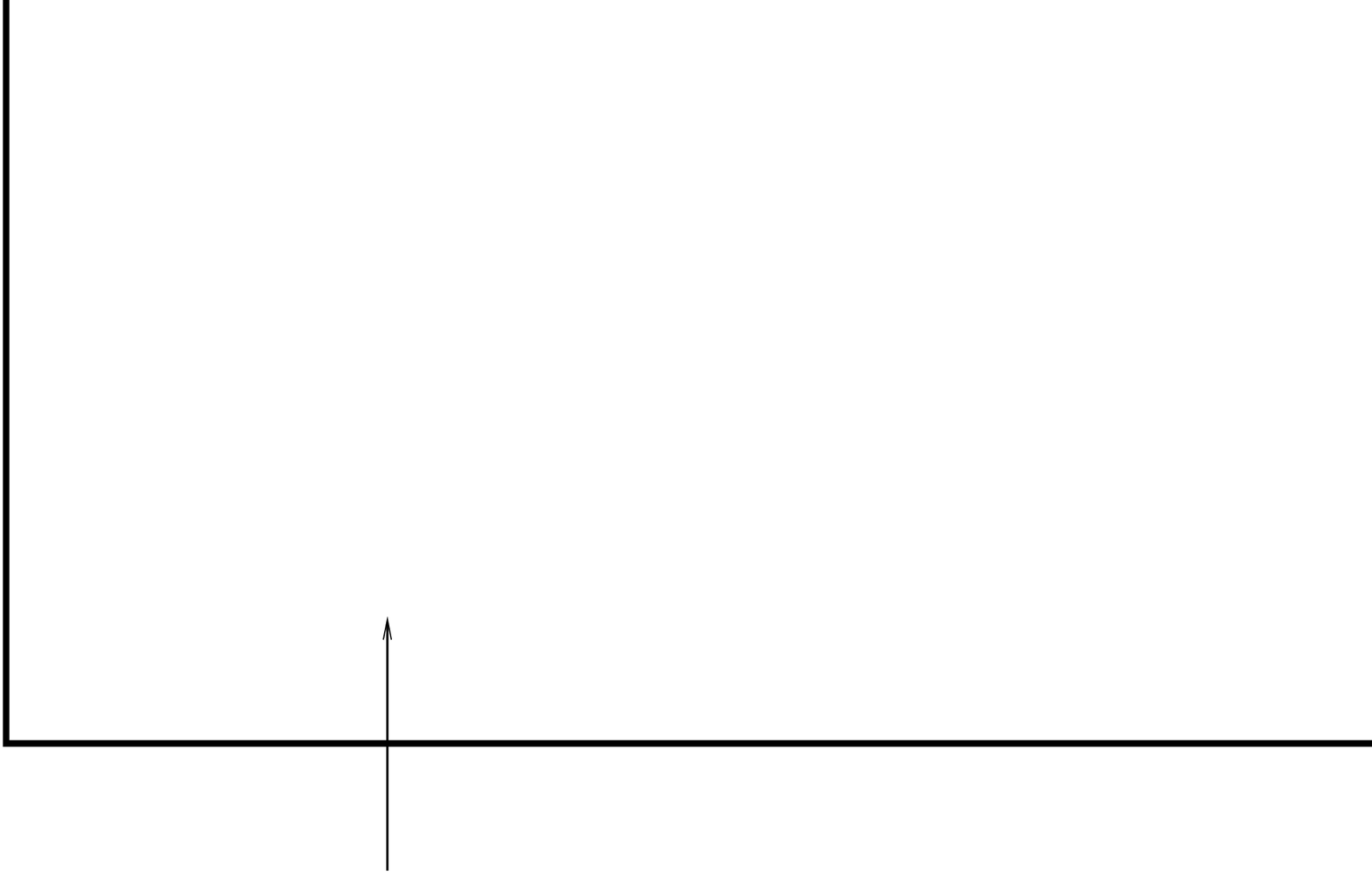}
\put(-335,81){\scriptsize$Q_{2}^{\mbox{-}2}$}
\put(-316,81){\scriptsize$\lambda_{1}\bar{Q}_{2}^{\mbox{-}2}$}
\put(-320,102){\scriptsize$\bar{Q}_{3+}^{0}$}
\put(-303,95){\scriptsize$Q_{3+}^{\mbox{-}2}$}
\put(-299,104){\scriptsize$\lambda_{2}$}
\put(-332,69.5){\scriptsize$\underline{Q}_{2}^{\mbox{-}3}$}
\put(-316,69.5){\scriptsize$\lambda_{1}Q_{2}^{\mbox{-}3}$}
\put(-323,50.5){\scriptsize$\bar{Q}_{3+}^{0}$}
\put(-305,53){\scriptsize$Q_{3+}^{\mbox{-}1}$}
\put(-300.5,44.5){\scriptsize$\lambda_{2}$}
\put(-293,82){\scriptsize$\underline{Q}_{1}^{4}$}
\put(-282,93){\scriptsize$Q_{1}^{3}$}
\put(-278,85){\scriptsize$\lambda_{3}$}
\put(-289.5,69){\scriptsize$Q_{1}^{\mbox{-}2}$}
\put(-276.5,62){\scriptsize$\bar{Q}_{1}^{\mbox{-}3}$}
\put(-281,55){\scriptsize$\lambda_{3}$}
\put(-251.5,92){\scriptsize$\bar{Q}_{2}^{0}$}
\put(-272,103){\scriptsize$Q_{2}^{\mbox{-}1}$}
\put(-267,94.5){\scriptsize$\lambda_{3}$}
\put(-248.5,59){\scriptsize$\bar{Q}_{2}^{0}$}
\put(-271,46){\scriptsize$Q_{2}^{\mbox{-}1}$}
\put(-266,55){\scriptsize$\lambda_{3}$}
\put(-227,71){\scriptsize$Q_{3\mbox{-}}^{\mbox{-}2}$}
\put(-258,72){\scriptsize$\lambda_{1}\bar{Q}_{3\mbox{-}}^{\mbox{-}2}$}
\put(-213,84.5){\scriptsize$\bar{Q}_{1}^{1}$}
\put(-240,84.5){\scriptsize$\lambda_{2}Q_{1}^{1}$}
\put(-119,81){\scriptsize$\underline{Q}_{1}^{3}$}
\put(-101.5,81){\scriptsize$\lambda_{2}Q_{1}^{3}$}
\put(-107,103){\scriptsize$\bar{Q}_{\hspace{-1pt}3+}^{\mbox{-}1}$}
\put(-88,100){\scriptsize$Q_{3+}^{\mbox{-}1}$}
\put(-87,108){\scriptsize$\lambda_{1}$}
\put(-119,69){\scriptsize$Q_{1}^{2}$}
\put(-101,69){\scriptsize$\lambda_{2}\bar{Q}_{1}^{2}$}
\put(-104.5,54){\scriptsize$\bar{Q}_{3+}^{\mbox{-}1}$}
\put(-87.5,56){\scriptsize$Q_{3+}^{\mbox{-}1}$}
\put(-83.5,47.5){\scriptsize$\lambda_{1}$}
\put(-76,82){\scriptsize$Q_{2}^{2}$}
\put(-64,94){\scriptsize$\bar{Q}_{2}^{3}$}
\put(-58,85.5){\scriptsize$\lambda_{3}$}
\put(-79,70){\scriptsize$\underline{Q}_{2}^{\mbox{-}4}$}
\put(-66.5,58.5){\scriptsize$Q_{2}^{\mbox{-}3}$}
\put(-59,67.5){\scriptsize$\lambda_{3}$}
\put(-35,91.5){\scriptsize$\bar{Q}_{1}^{0}$}
\put(-52.5,104){\scriptsize$Q_{1}^{1}$}
\put(-51.5,95.5){\scriptsize$\lambda_{3}$}
\put(-35.5,56){\scriptsize$\bar{Q}_{1}^{0}$}
\put(-50.5,47){\scriptsize$Q_{1}^{1}$}
\put(-52.5,55){\scriptsize$\lambda_{3}$}
\put(-13.5,71){\scriptsize$Q_{3\mbox{-}}^{2}$}
\put(-42,71){\scriptsize$\lambda_{1}\bar{Q}_{3\mbox{-}}^{3}$}
\put(0,84){\scriptsize$\bar{Q}_{2}^{\mbox{-}1}$}
\put(-26.5,83.5){\scriptsize$Q_{2}^{\mbox{-}1}\hspace{-1pt}\lambda_{2}$}
\put(-364,111){$N_{1}$} \put(-367,64){$[\lambda_{3}]$}
\put(-364,18){$N_{2}$}
\put(-148,111){$N_{1}$} \put(-151,64){$[\lambda_{3}]$}
\put(-148,18){$N_{2}$}
}
\caption{A more accurate depiction of fig.\
\hspace{-3pt}\ref{sweepnf1a} for a particular
$n$}\label{sweepnf1b}}

The gauge bosons produced become unstable, as explained in
\cite{me}, on curves encircling the cores. If we define
$\bar{S}$'s in the same way as (\ref{sss}) but for $\bar{Q}$'s,
then we have within one section of core only the states (fixed
$n$, all $m$),  $Q_{i}^{n}, Q_{i}^{n+1}, \bar{Q}_{i}^{n},
\underline{Q}_{i}^{n+1}, S_{m,\, i}^{n-1}, S_{m,\, i}^{n},
\bar{S}_{m,\, i}^{n-1},\bar{S}_{m,\, i}^{n}$ and the three quarks.
A typical core is shown in fig.\ \hspace{-3pt}(\ref{corenf1}).

\FIGURE[h]{ \centering
\makebox[13cm]{\includegraphics[width=13cm]{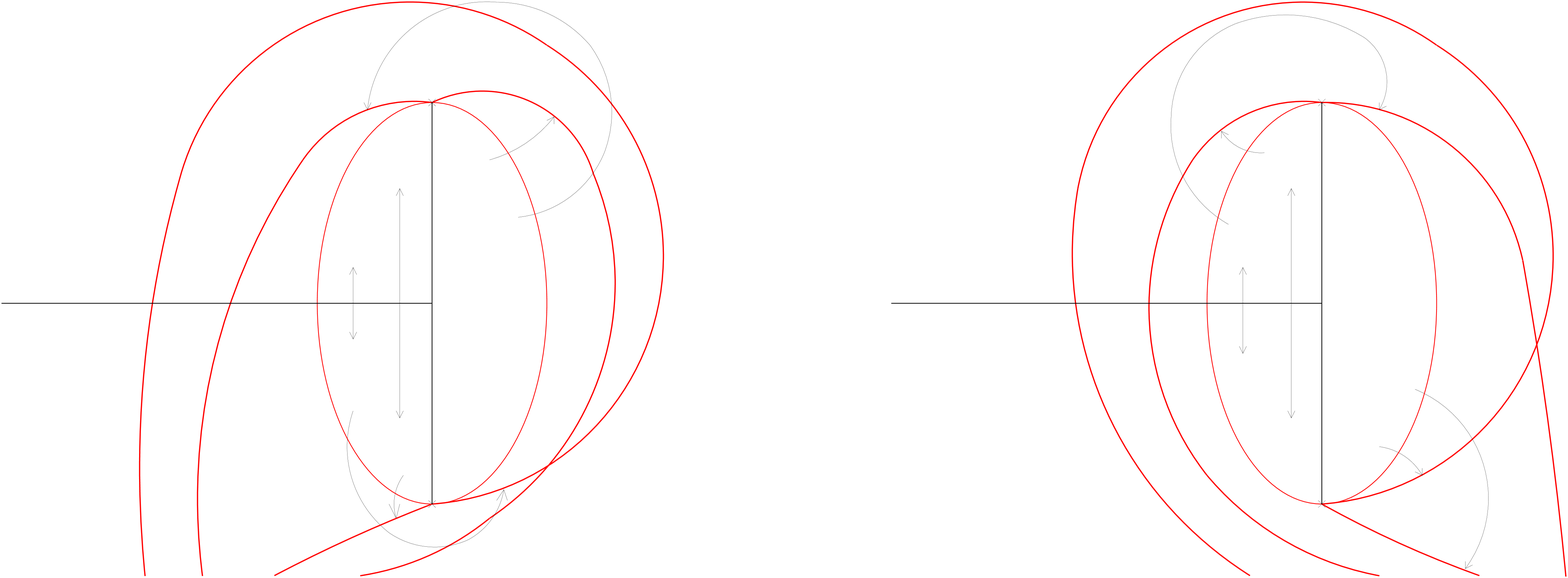}
\put(-280,96){\scriptsize$\bar{Q}_{1}^{0}$}
\put(-280,30){\scriptsize$\underline{Q}_{1}^{3}$}
\put(-290,80){\scriptsize$\underline{Q}_{1}^{1}$}
\put(-290,44){\scriptsize$\bar{Q}_{1}^{2}$}
\put(-264,96){\scriptsize$\underline{Q}_{1}^{2}$}
\put(-260,81){\scriptsize$\bar{Q}_{1}^{1}$}
\put(-70,96){\scriptsize$\bar{Q}_{1}^{0}$}
\put(-70,30){\scriptsize$\underline{Q}_{1}^{3}$}
\put(-80,80){\scriptsize$\underline{Q}_{1}^{1}$}
\put(-80,44){\scriptsize$\bar{Q}_{1}^{2}$}
\put(-54,29){\scriptsize$\underline{Q}_{1}^{2}$}
\put(-48,41){\scriptsize$\bar{Q}_{1}^{1}$}
\put(-335.5,10){\scriptsize$\underline{Q}_{1}^{4}\hspace{-4.5pt}\rightarrow
\hspace{-2pt}Q_{1}^{3}\lambda_{3}$}
\put(-26,10){\scriptsize$Q_{1}^{1}\lambda_{3}\hspace{-3pt}\leftarrow\hspace{-2pt}\bar{Q}_{0}^{1}$}
}\caption{The consistency of extra states in the $\alpha_{1}$
core}\label{corenf1s}}

\FIGURE[h]{ \centering
\makebox[6cm]{\includegraphics[width=6cm]{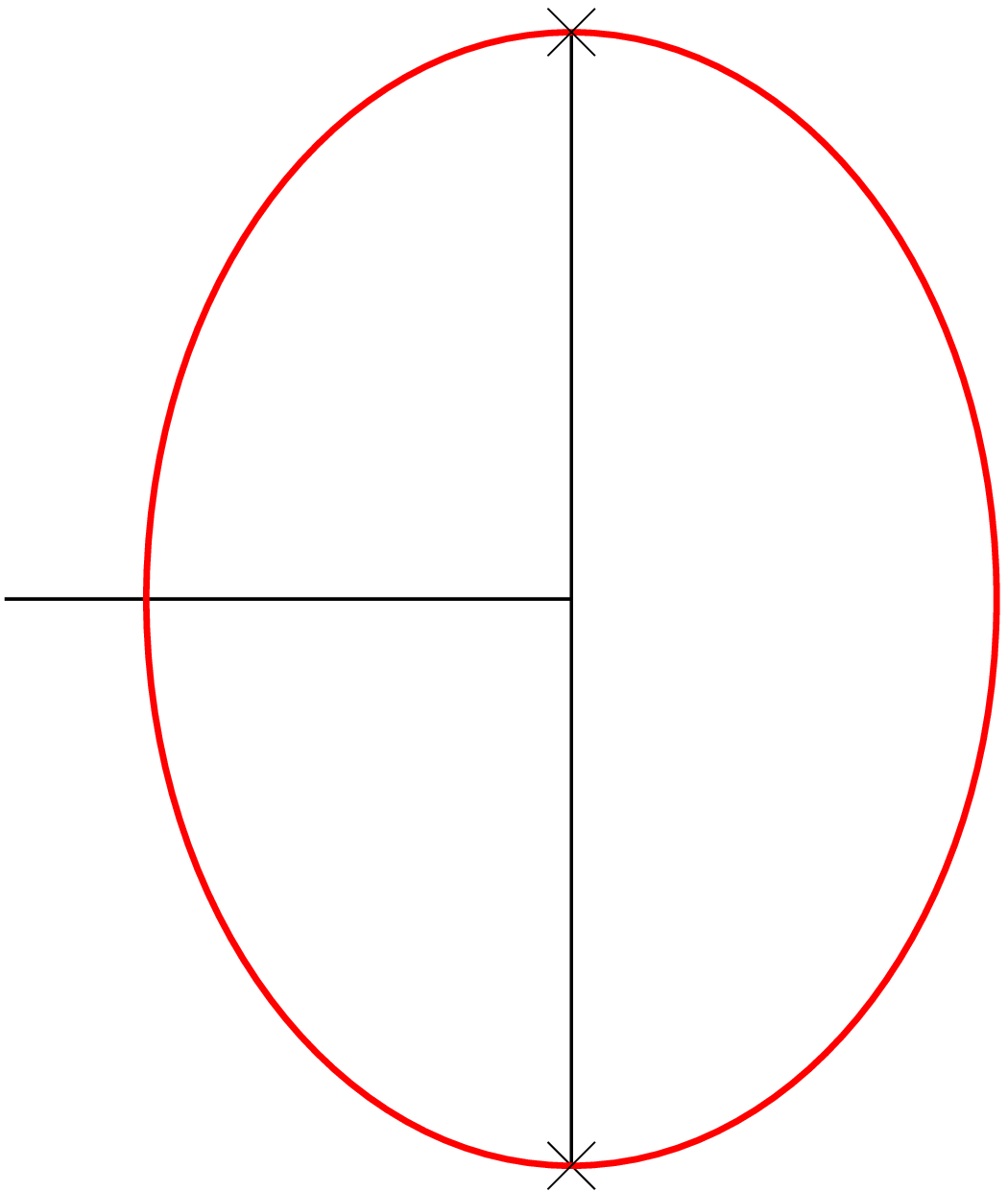}
\put(-86,178){\scriptsize$Q_{i}^{0}$}
\put(-103,165){\scriptsize$Q_{i}^{1}$}
\put(-95,150){\scriptsize$S_{m,i}^{\mbox{-}1}$}
\put(-95,120){\scriptsize$S_{m,i}^{0}$}
\put(-116,139){\scriptsize$\bar{S}_{m,i}^{\mbox{-}1}$}
\put(-116,109){\scriptsize$\bar{S}_{m,i}^{0}$}
\put(-131,140){\scriptsize$\lambda_{1}$}
\put(-131,125){\scriptsize$\lambda_{2}$}
\put(-131,110){\scriptsize$\lambda_{3}$}
\put(-115,165){\scriptsize$\underline{Q}_{i}^{1}$}
\put(-98,178){\scriptsize$\bar{Q}_{i}^{0}$}
\put(-86,19){\scriptsize$Q_{i}^{2}$}
\put(-103,32){\scriptsize$Q_{i}^{3}$}
\put(-95,47){\scriptsize$S_{m,i}^{1}$}
\put(-95,77){\scriptsize$S_{m,i}^{2}$}
\put(-116,58){\scriptsize$\bar{S}_{m,i}^{1}$}
\put(-116,88){\scriptsize$\bar{S}_{m,i}^{2}$}
\put(-131,57){\scriptsize$\lambda_{1}$}
\put(-131,72){\scriptsize$\lambda_{2}$}
\put(-131,87){\scriptsize$\lambda_{3}$}
\put(-115,32){\scriptsize$\underline{Q}_{i}^{3}$}
\put(-98,19){\scriptsize$\bar{Q}_{i}^{2}$}
\put(-50,40){\scriptsize$Q_{i}^{1}$}
\put(-50,53){\scriptsize$Q_{i}^{2}$}
\put(-45,86){\scriptsize$S_{m,i}^{0}$}
\put(-45,111){\scriptsize$S_{m,i}^{1}$}
\put(-66,77){\scriptsize$\bar{S}_{m,i}^{0}$}
\put(-66,121){\scriptsize$\bar{S}_{m,i}^{1}$}
\put(-21,85){\scriptsize$\lambda_{1}$}
\put(-21,100){\scriptsize$\lambda_{2}$}
\put(-21,115){\scriptsize$\lambda_{3}$}
\put(-50,140){\scriptsize$\underline{Q}_{i}^{2}$}
\put(-50,153){\scriptsize$\bar{Q}_{i}^{1}$}
}\caption{The states within the single core for one flavour, large
$R/\Lambda$}\label{corenf1}}

Double cores, where they might begin to exist here, are also
highly similar to the pure case. Additional curves surrounding
them signal the place where the $\bar{S}$'s with large $m$, like
their unbarred counterparts, are no longer stable. On each side of
the branch cut $V$ we have four $S$'s, corresponding to them, four
(of each flavour)  $\bar{S}$'s, $N_{f}$ flavours of three quarks,
as well as the two states which become massless in that region of
that arm's core connected to this side of $V$. An example can be
found in fig.\ \hspace{-3pt}(\ref{dcorenf1a}).

\FIGURE[t]{ \centering
\makebox[13cm]{\includegraphics[width=14cm,height=7cm]{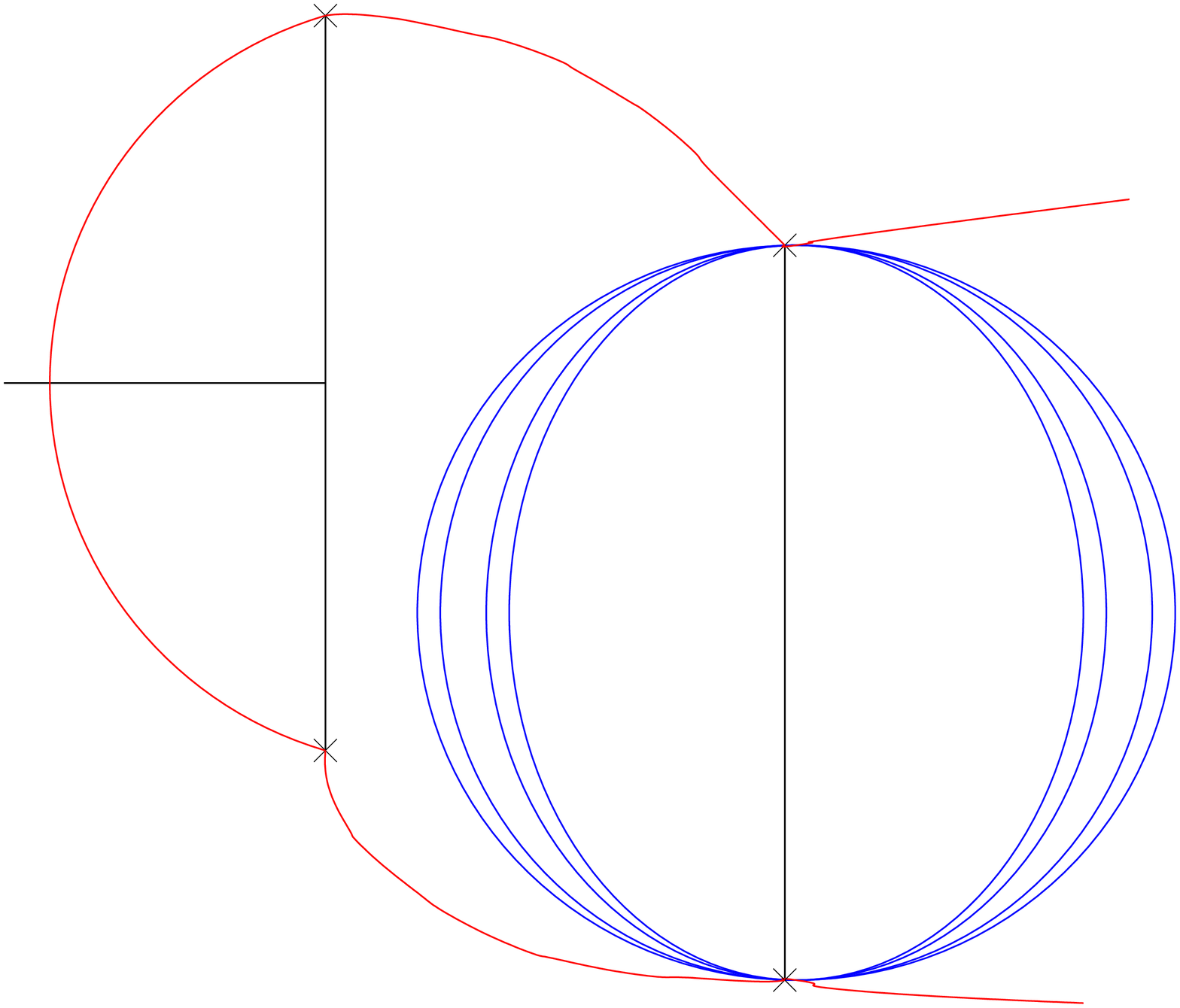}
\put(-280,170){$\bar{Q}_{1}^{1}$}
\put(-265,170){$\underline{Q}_{1}^{2}$}\put(-250,170){$Q_{1}^{1}$}
\put(-235,170){$Q_{1}^{2}$}
\put(-280,150){$S_{m,1}^{1}$} \put(-260,150){$S_{m,1}^{0}$}
\put(-280,130){$\bar{S}_{m,1}^{1}$}
\put(-260,130){$\bar{S}_{m,1}^{0}$}
\put(-275,110){$\lambda_{1}$} \put(-275,95){$\lambda_{2}$}
\put(-275,80){$\lambda_{3}$}
\put(-169,120){$Q_{1}^{2}$} \put(-169,105){$Q_{1}^{1}$}
\put(-169,45){$Q_{2}^{0}$} \put(-169,30){$Q_{2}^{1}$}
\put(-179,85){$Q_{3+}^{1}$} \put(-179,70){$Q_{3+}^{2}$}
\put(-114,120){ $Q_{2}^{1}$} \put(-114,105){ $Q_{2}^{0}$}
\put(-114,45){$Q_{1}^{1}$} \put(-114,30){$Q_{1}^{2}$}
\put(-104,85){$Q_{3-}^{2}$} \put(-104,70){$Q_{3-}^{3}$}
\put(-199,55){$\bar{Q}_{2}^{0}$} \put(-199,40){$\bar{Q}_{2}^{1}$}
\put(-209,95){$\bar{Q}_{3+}^{1}$}
\put(-209,80){$\bar{Q}_{3+}^{2}$}
\put(-84,55){$\bar{Q}_{1}^{1}$} \put(-84,40){$\bar{Q}_{1}^{2}$}
\put(-74,95){$\bar{Q}_{3-}^{2}$} \put(-74,80){$\bar{Q}_{3-}^{3}$}
\put(-150,95){$\lambda_{1}$} \put(-150,80){$\lambda_{2}$}
\put(-150,65){$\lambda_{3}$}
\put(-126,95){$\lambda_{1}$} \put(-126,80){$\lambda_{2}$}
\put(-126,65){$\lambda_{3}$}
}\caption{The states within the double core for one
flavour}\label{dcorenf1a}}

\subsection{Larger $\Lambda$}

Considering the moduli space at a fixed radius, we find there are
three regimes, each more complex than the next. First, the
infinite radius limit. Here the cuts and curves for all flavours
with any bare masses are in identical positions. Shrinking $R$,
but keeping $R\gg \Lambda$, $R\gg m_{i}$, we get models which are
largely the same, the differences following a simple pattern, such
as having a monodromy which acts, after three rotations, as
$n\rightarrow n-(6-N_{f})$. They contain simple cores where one of
the U$(1)$'s becomes strongly coupled. Finally, we get to $R\sim
\Lambda$. Even for SU$(2)$ this is not formulaic, but almost every
case needs to be treated individually, having its own accidental
symmetries and other quirks.

For SU$(3)$, the job of pinning everything down exactly becomes
unwieldy. We have limited our scope to just the generics. We can
determine, for instance, how many double cores there will be, and
roughly where they lie, for the massless and the equal bare mass
cases. Except in the simplest examples, understanding exactly how
the trefoliate curve degenerates into circles, and precisely the
intermediate monodromies and states is a weighty challenge which
we have not properly met. The one massless flavour example is a
case in point: Its particular foible, as one can see from Table
\ref{tab2} in the Appendix, is that the quark singularity does not
factor off as in the other massless cases, except in the large
$R/\Lambda$ limit. For merely large $R/\Lambda$ we will see it in
our three-spherical sections as portrayed above, a separate
circular singularity. When $R\sim \Lambda$, however, it must merge
with the remains of the other, what was trefoliate, curve. At a
radius where this has begun to occur, from the perspective of our
radial sections, this can only happen when the quark singularity
and part of the old trefoil {\em with exactly the same monodromy}
coincide, then disappear from view. We can thus deduce that the
quark monodromy must pass through a core in order to become
dyonic. This is what happens in general in SU$(2)$ to the
singularity of a massive quark when we reduce $|m|$ to be very
small.

In order to get to a position which does this, it is obvious that
(as we shrink $R$) the quark singularity must (repeatedly)
intersect the other curve before passing through it. From table
\ref{tab2}, we can see that, if $v$ is roughly zero when this
happens (the quark singularity only picks up corrections at order
$(\Lambda^{5})^{3}$) then there are five intersections. This fills
a faithful representation of the five-fold symmetry listed in
Table \ref{tab0}. An example of where we believe this merging
occurs is shown in fig.\ \hspace{-3pt}(\ref{bqnf1}).

\FIGURE[t]{ \centering
\makebox[10cm]{\includegraphics[width=10cm]{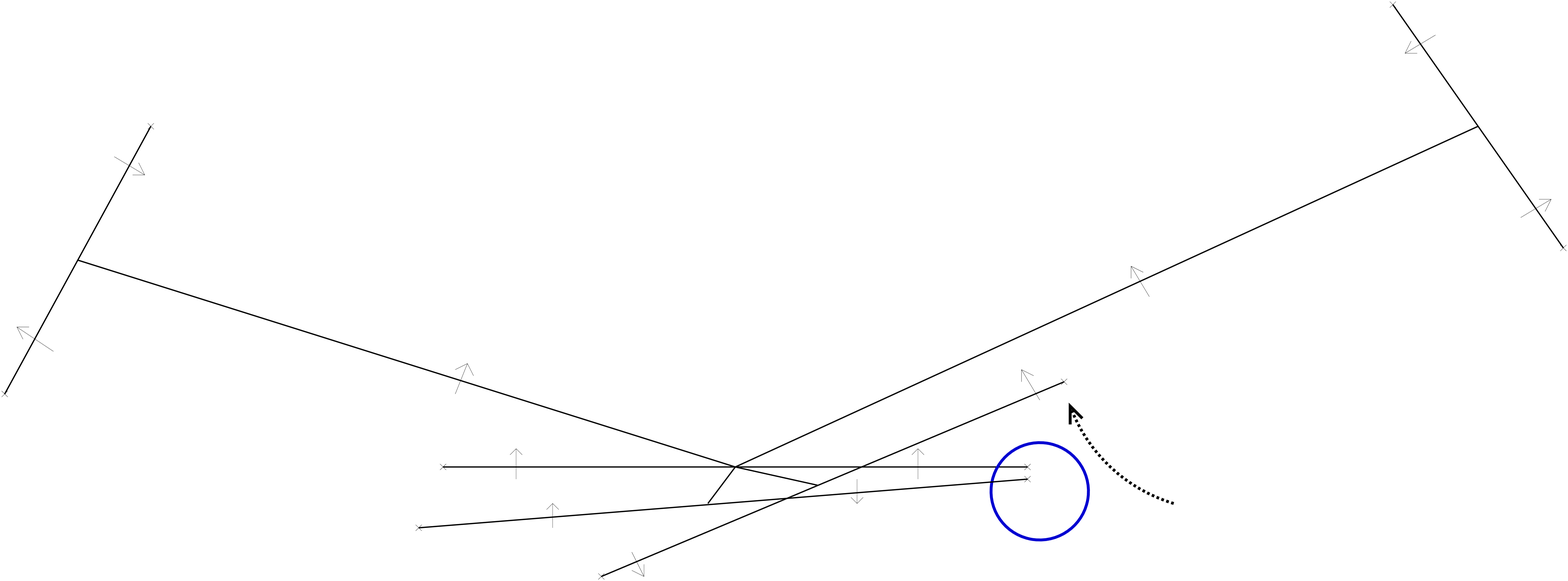}
\put(-139,5){\scriptsize$[\bar{Q}_{3}^{0}]$}
\put(-124,9.5){\scriptsize$[\bar{Q}_{3}^{0}]$}
\put(-80,45){\scriptsize$N_{1}$} \put(-203,43){\scriptsize$N_{2}$}
}\caption{A place in the moduli space (ringed) where what was the
quark singularity is equal and opposite, and nearly superposed
with a dyonic singularity prior to merging.}\label{bqnf1}}

If we continue to reduce $R/\Lambda$ we believe something akin to
the quark singularity then re-emerges, we then get to a value
(which can be read from Table \ref{tab3} in the Appendix) where
the trefoliate curve self-intersects and double cores arise, one
at each of five points. In the singularity/core crossings too
there will be, as there must, accumulation points, and some curves
following into the core, joining these accumulation points to what
begins as the quark singularity. These are a reduced set of those
curves previously attached to the quark singularity, the rest the
accumulation points restrict to end temporarily outside the core.
Those inside involve only the states present within the core and
their values after traversing the quark monodromy. Fig.\
\hspace{-3pt}(\ref{bqcorenf1}) shows a sketch of this.

\FIGURE[t]{ \centering
\makebox[10cm]{\includegraphics[width=8cm]{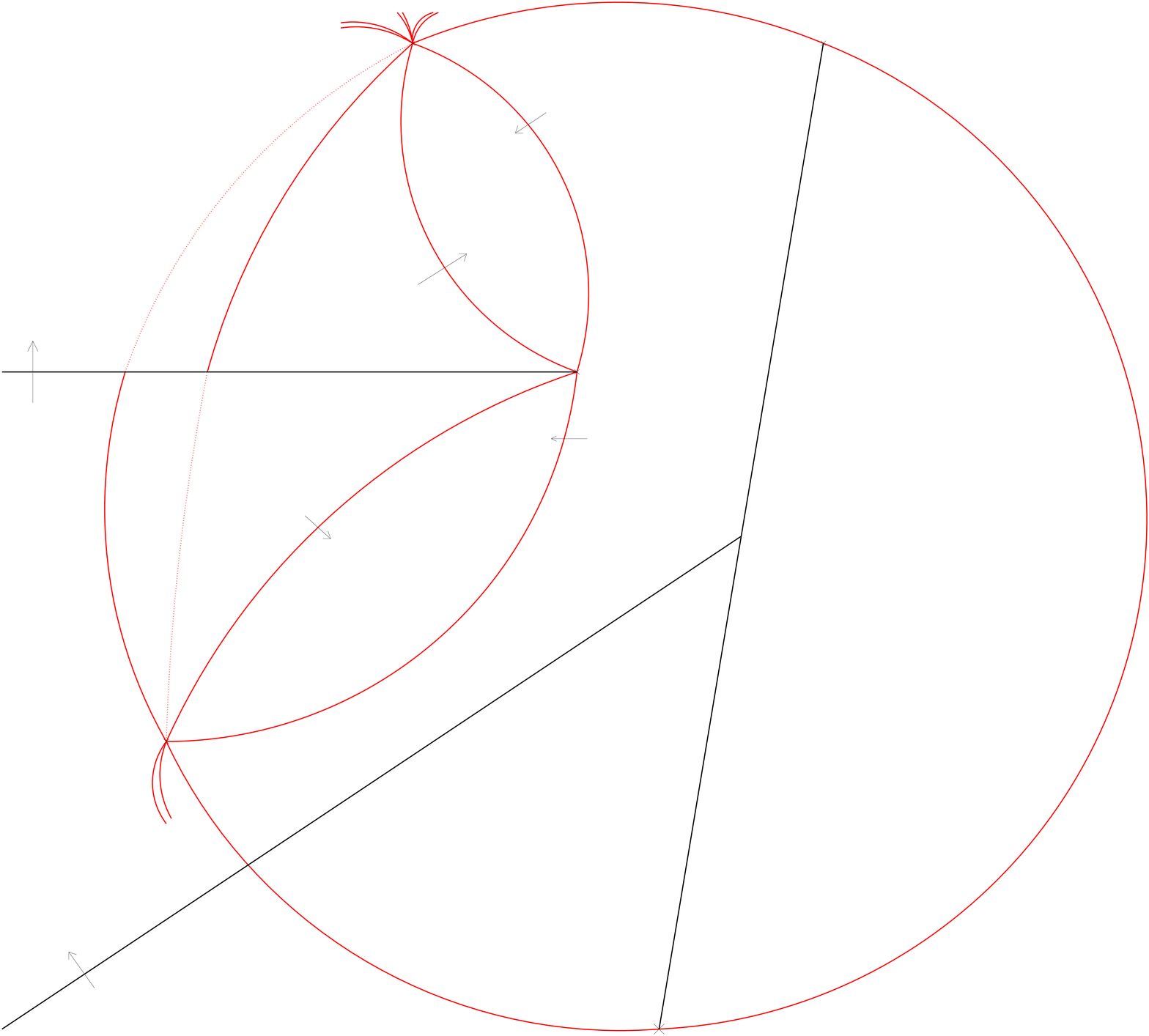}
\put(-231,114){$[\lambda_{3}]$} \put(-213,0){$N_{2}$}
\put(-153,144){$\underline{1}$}\put(-132,155){$1$}
\put(-134,173){$1$}\put(-116,181){$\bar{1}$}
\put(-127,115){$2$}\put(-109,115){$\bar{2}$}
\put(-174,106){$\underline{2}$}\put(-159,92){$2$}
}\caption{A quark singularity intersecting a single core. Note the
formation of accumulation points and curves inwards from them. A
$\bar{2}$ is any state with magnetic charge 2, {\it etc.}. There
will be a curve for each.}\label{bqcorenf1}}

If we add mass $|m|\ll \Lambda$ then we see little difference,
even at $R\sim |m|$, as by the time we have shrunk $R\ll \Lambda$
the curve has disappeared anyway. For more flavours, we do see a
qualitative difference, however. The quark singularity, which we
can treat separately from the rest, keeps its integrity as a
circle all the way down to $R=0$, but when we add mass it shrinks
to a point before we reduce $R$ all the way.

\section{More Flavours}

We shall pick the case with three flavours (usually with nearly
equal masses) to show how the previous argument generalises. Let
us begin again with the simplest region at $R/\Lambda=\infty$.
This time the quark monodromy becomes $[(0,\lambda_{3})_{100}]
[(0,\lambda_{3})_{010}][(0,\lambda_{3})_{001}]$, or
$[\lambda^{a}_{3}][\lambda^{b}_{3}][\lambda^{c}_{3}]$ in
simplified notation. The entire set of $Q_{3}$'s is invariant
under this. It adds three upper bars to a $Q_{2}$ (cancelling with
any lower ones), and adds three lower bars to the $Q_{1}$'s. Thus
we see that $Q_{2}$ up to $\quuu_{2}$ \mbox{} exist above the cut,
and transform to $Q_{2}$ down to $\qddd_{2}$ \mbox{} below. The
reverse is true of the $Q_{1}$'s with the lower bars existing
above the cut in the $3+$ region. Lower-barred $Q_{3}$'s do not
exist anywhere, and all the upper-barred $Q_{3\pm}$'s exist in the
same region as the pure ones.

If we restrict ourselves to large $R/\Lambda$, as already
mentioned, the picture changes very little. The $G$ curves, which
partition the $Q_{3}$'s, rotate to generate, not $\bar{G}$'s, but
$\guuu$'s for three flavours, {\it etc.\ }\hspace{-3pt}, where all
three bars persist on the component of the bound state not
becoming massless. This is natural because, of course, in this
sector only unbarred states become massless.

Now we need to find the curves which partition all the
intermediate $Q_{3\pm}$'s as well as all the lower-barred and
upper-barred $Q_{1}$'s in the $3+$(upper) and $3-$ regions
respectively; and similarly, the lower-barred and upper-barred
$Q_{2}$'s in $3+$(lower) and $3-$. These are just like the second
batch of global curves described for one flavour. There are
$N_{f}2^{N_{f}-1}$ copies of each of the two curves depicted in
fig.\ \hspace{-3pt}(\ref{sweepnf1b}). When all the bare masses are
equal, the curves superpose. If not, they come in $N_{f}$ layers.
Everywhere where a barred state becomes marginally stable to a
pure state in the one flavour case, now a double barred will decay
to the single barred one (outside it, if the masses differ); and a
triple-barred one to the double, and so on up to $N_{f}$. Also in
the one flavour case were places on these curves where a pure
state would have the mass of a barred one plus a quark, for
instance, $Q_{2}\rightarrow \bar{Q}_{2}+\lambda_{3}$. As the quark
monodromy now changes states by $N_{f}$ bars, we find it becomes
the second highest number of bars going to the highest, {\it e.g.\
}\hspace{-3pt} for three flavours, equal masses,
$\quu_{2}\rightarrow\quuu_{2}+\lambda_{3}$. Superposed are other
curves we might think of as forming a chain
\[Q_{2}\rightarrow\lambda_{3}+\bar{Q}_{2}\rightarrow\lambda_{3}+\quu_{2}
\rightarrow\lambda_{3}+\quuu_{2} .\] We show one of the two curves
in detail in fig.\ \hspace{-3pt}(\ref{sweepnf3a}).

\FIGURE[t]{ \centering
\makebox[13cm]{\includegraphics[width=13cm,height=19cm]{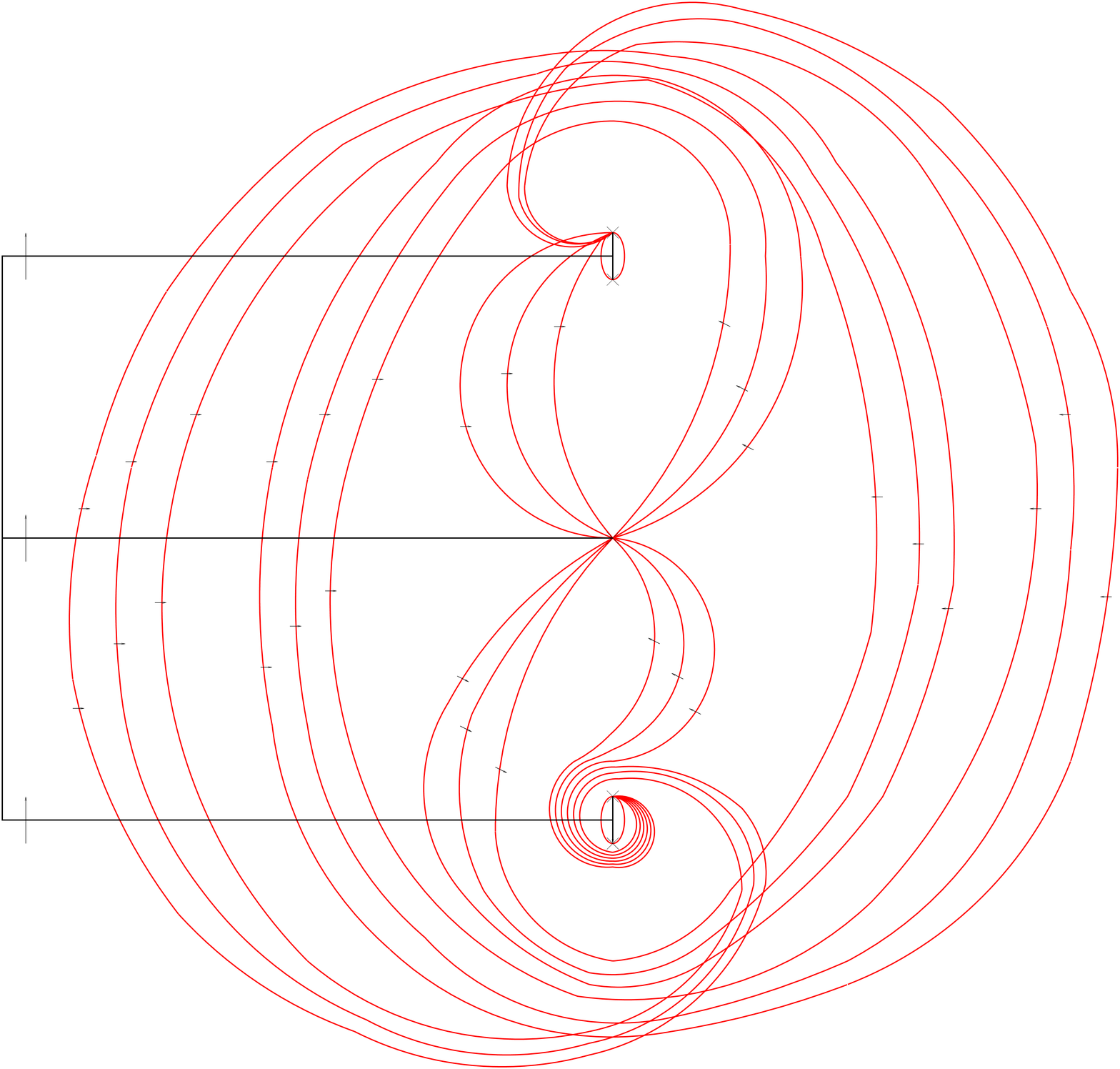}
\put(-367,390){$N_{1}$} \put(-368,256){$[\lambda_{3}]^{3}$}
\put(-367,107){$N_{2}$}
\put(-229,324){\scriptsize$\underline{\underline{\underline{Q}}}_{\raisebox{5pt}{\tiny
1}}^{6}$}
\put(-213,324){\scriptsize$\underline{\underline{Q}}_{\raisebox{3pt}{\tiny
1}}^{5bc}\lambda_{3}^{a}$}
\put(-217,351){\scriptsize$\underline{\underline{Q}}_{\raisebox{3pt}{\tiny
1}}^{5bc}$}
\put(-200,351){\scriptsize$\underline{Q}_{1}^{4c}\lambda_{3}^{b}$}
\put(-198,374){\scriptsize$\underline{Q}_{1}^{4c}$}
\put(-183,374){\scriptsize$Q_{1}^{3}\lambda_{3}^{c}$}
\put(-321,329){\scriptsize$\bar{\bar{Q}}_{2}^{0ab}$}
\put(-303,329){\scriptsize$\bar{\bar{\bar{Q}}}_{2}^{0}\lambda_{1}^{c}$}
\put(-344,304){\scriptsize$\bar{Q}_{2}^{0a}$}
\put(-323,304){\scriptsize$\bar{\bar{Q}}_{2}^{0ab}\lambda_{1}^{b}$}
\put(-356,280){\scriptsize$Q_{2}^{0}$}
\put(-340,280){\scriptsize$\bar{Q}_{2}^{0a}\lambda_{1}^{a}$}
\put(-332,234){\scriptsize$\underline{Q}_{2}^{\mbox{-}3c}$}
\put(-314,234){\scriptsize$Q_{2}^{\mbox{-}1}\lambda_{1}^{c}$}
\put(-349,212.5){\scriptsize$\underline{\underline{Q}}_{\raisebox{3pt}{\tiny
2}}^{\mbox{-}3bc}$}
\put(-328,212.5){\scriptsize$\underline{Q}_{2}^{\mbox{-}3c}\lambda_{1}^{b}$}
\put(-360,179){\scriptsize$\underline{\underline{\underline{Q}}}_{\raisebox{5pt}{\tiny
2}}^{\mbox{-}3}$}
\put(-342,179){\scriptsize$\underline{\underline{Q}}_{\raisebox{3pt}{\tiny
2}}^{\mbox{-}3bc}\lambda_{1}^{a}$}
\put(-261,347){\scriptsize$\bar{Q}_{3+}^{\mbox{-}1c}$}
\put(-243,346.5){\scriptsize$Q_{3+}^{\mbox{-}1}\hspace{-1pt}\lambda_{2}^{c}$}
\put(-278,328){\scriptsize$\bar{\bar{Q}}_{3+}^{\mbox{-}1bc}$}
\put(-259,328){\scriptsize$\bar{Q}_{3+}^{\mbox{-}1c}\lambda_{2}^{b}$}
\put(-295,304.5){\scriptsize$\bar{\bar{\bar{Q}}}_{3+}^{\mbox{-}1}$}
\put(-276,304.5){\scriptsize$\bar{\bar{Q}}_{3+}^{\mbox{-}1bc}\lambda_{2}^{a}$}
\put(-277,241){\scriptsize$\bar{Q}_{3+}^{\mbox{-}1c}$}
\put(-258,240){\scriptsize$Q_{3+}^{\mbox{-}1}\lambda_{2}^{c}$}
\put(-289,222){\scriptsize$\bar{\bar{Q}}_{3+}^{\mbox{-}1bc}$}
\put(-268,222){\scriptsize$\bar{Q}_{3+}^{\mbox{-}1c}\lambda_{2}^{b}$}
\put(-300,199){\scriptsize$\bar{\bar{\bar{Q}}}_{3+}^{\mbox{-}1}$}
\put(-281,199){\scriptsize$\bar{\bar{Q}}_{3+}^{\mbox{-}1bc}\lambda_{2}^{a}$}
\put(-216,149){\scriptsize$\bar{\bar{Q}}_{1}^{\mbox{-}2ab}$}
\put(-203,139){\scriptsize$\bar{\bar{\bar{Q}}}_{1}^{\mbox{-}3}$}
\put(-203,129){\scriptsize$\lambda_{3}^{c}$}
\put(-233,171){\scriptsize$\bar{Q}_{1}^{\mbox{-}1a}$}
\put(-214,165){\scriptsize$\bar{\bar{Q}}_{1}^{\mbox{-}2ab}\lambda_{3}^{b}$}
\put(-231,198){\scriptsize$Q_{1}^{0}$}
\put(-218,187){\scriptsize$\bar{Q}_{1}^{\mbox{-}1a}\lambda_{3}^{a}$}
\put(-121,303){\scriptsize$\bar{\bar{\bar{Q}}}_{2}^{2}$}
\put(-146,313){\scriptsize$\lambda_{3}^{a}\bar{\bar{Q}}_{2}^{1bc}$}
\put(-124,333){\scriptsize$\bar{\bar{Q}}_{2}^{1bc}$}
\put(-148,344){\scriptsize$\lambda_{3}^{b}\bar{Q}_{2}^{0c}$}
\put(-129,370.5){\scriptsize$\bar{Q}_{2}^{0c}$}
\put(-153,377){\scriptsize$\lambda_{3}^{c}Q_{2}^{\mbox{-}1}$}
\put(-136.5,174){\scriptsize$\bar{\bar{\bar{Q}}}_{2}^{2}$}
\put(-163,181){\scriptsize$\lambda_{3}^{a}\bar{\bar{Q}}_{2}^{1bc}$}
\put(-143,193){\scriptsize$\bar{\bar{Q}}_{2}^{1bc}$}
\put(-171,198){\scriptsize$\lambda_{3}^{b}\bar{Q}_{2}^{0c}$}
\put(-152,210){\scriptsize$\bar{Q}_{2}^{0c}$}
\put(-176,216){\scriptsize$\lambda_{3}^{c}Q_{2}^{\mbox{-}1}$}
\put(0,236){\scriptsize$\bar{\bar{\bar{Q}}}_{1}^{1}$}
\put(-30,236){\scriptsize$\lambda_{2}^{a}\bar{\bar{Q}}_{1}^{1bc}$}
\put(-14,328){\scriptsize$\bar{\bar{Q}}_{1}^{1bc}$}
\put(-41,329){\scriptsize$\lambda_{2}^{b}\bar{Q}_{1}^{1c}$}
\put(-25,280){\scriptsize$\bar{Q}_{1}^{1c}$}
\put(-49,280){\scriptsize$\lambda_{2}^{c}Q_{1}^{1}$}
\put(-53,230){\scriptsize$Q_{3\mbox{-}}^{0}$}
\put(-83,230){\scriptsize$\lambda_{1}^{a}\bar{Q}_{3\mbox{-}}^{0a}$}
\put(-63,263){\scriptsize$\bar{Q}_{3\mbox{-}}^{0a}$}
\put(-89,263){\scriptsize$\lambda_{1}^{b}\bar{\bar{Q}}_{3\mbox{-}}^{0ab}$}
\put(-77,287){\scriptsize$\bar{\bar{Q}}_{3\mbox{-}}^{0ab}$}
\put(-102,287){\scriptsize$\lambda_{1}^{c}\bar{\bar{\bar{Q}}}_{3\mbox{-}}^{0}$}
}\caption{A look at one chain from one of the two sets of curves
that help partition the three regions for $N_{f}=3$
}\label{sweepnf3a}}

When the masses are not equal then we have to take into account
the s-charges. For three flavours, single barred states will have
one index, equal to $a, b,$ or $c$, double bars will have two
indices, which cannot be equal, and we can ignore these for
triples which must have one of each. A typical equivalent of the
previous chain would be
\[Q_{2}\rightarrow\lambda^{a}_{3}+\bar{Q}^{a}_{2}\rightarrow\lambda^{c}_{3}+\quu^{ac}_{2}
\rightarrow\lambda^{b}_{3}+\quuu_{2} .\] Including all six
permutations of the indices gives 12 curves. Due to the
symmetries, there will take 3 of each curve to cover all $n$.
There would be another 36 such curves related to the second
diagram in fig.\ \hspace{-3pt}(\ref{sweepnf1b}). For $N_{f}=1$ to
5 we have $5, 16, 36, 64$ and 80 examples respectively of each of
the two types of curve.

A single core will contain the same states as the pure core plus
whatever other states are left in that vicinity after all the
previous curves have reduced the weak-coupling spectrum. We know
this to be only the largest upper bars of the states not related
to the core, {\it e.g.\ }\hspace{-3pt} for the $\alpha_{1}$ core
with three flavours, each region of the core contains
$S_{m,1}^{n}$ for all $m$ and two consecutive $n$ (let us take
$n=0,1$), as well as the same
${\raisebox{2pt}{$\bar{\hspace{9.5pt}}$}\hspace{-9pt}\raisebox{3.5pt}{$\bar{\hspace{9.5pt}}$}\hspace{-9.5pt}
\bar{S}}_{m,1}^{n}$'s. It contains the quarks
$\lambda_{i}^{a},\lambda_{i}^{b}$ and $\lambda_{i}^{c}$ for all
three $i$. It also contains the maps of these through the quantum
cut in the core, namely $\bar{Q}_{1}^{1}$ and
$\underline{Q}_{1}^{2}$ for all s-charges. Finally, it contains
the SU$(2)$ embeddings $Q_{1}^{1}$ and $Q_{1}^{2}$.

A double core, similarly, on the side touching the above single
core will contain all ($N_{c}N_{f}=$) nine quarks, plus
$Q_{1}^{1}$ and $Q_{1}^{2}$. Also $Q_{-2,\, 1}^{-2}$ and $Q_{-2,\,
1}^{-1}$, and $Q_{2,\, 1}^{6}$ and $Q_{2,\, 1}^{7}$; and the most
upper-barred version of these last four, $\quuu_{-2,\, 1}^{-2},\,
\quuu_{-2,\, 1}^{-1},\, \quuu_{2,\, 1}^{6}$, and $\quuu_{2,\,
1}^{7}$. (The dyons produced by the quarks traversing the quantum
cut lose stability on curves lying between the accumulation points
too).

To complete the picture, we show what happens when a quark
singularity, {\it e.g.\ }\hspace{-3pt}
$[\lambda^{a}_{3}][\lambda^{b}_{3}][\lambda_{3}^{c}]$ crosses a
double core region as above. The number of states remains the
same, just the transformation that would have occurred had the
complete double core and its adjoining single cores passed through
the quark singularity is enacted by CMS's connecting the
singularity with accumulation points on the boundary of the core,
sweeping across the intervening space. A sketch of this is shown
in fig.\ \hspace{-3pt}(\ref{dcorenf3b}). Note that although no
intermediate barred states are stable in the core for equal masses
(they are marginally stable on the sweeping CMS's), when we
perturb this situation a little by adding tiny mass differences
and the quark singularity splits a little, such states can exist
between the newly distinct sweeping curves.

\FIGURE[h]{ \centering
\makebox[10cm]{\includegraphics[width=8cm]{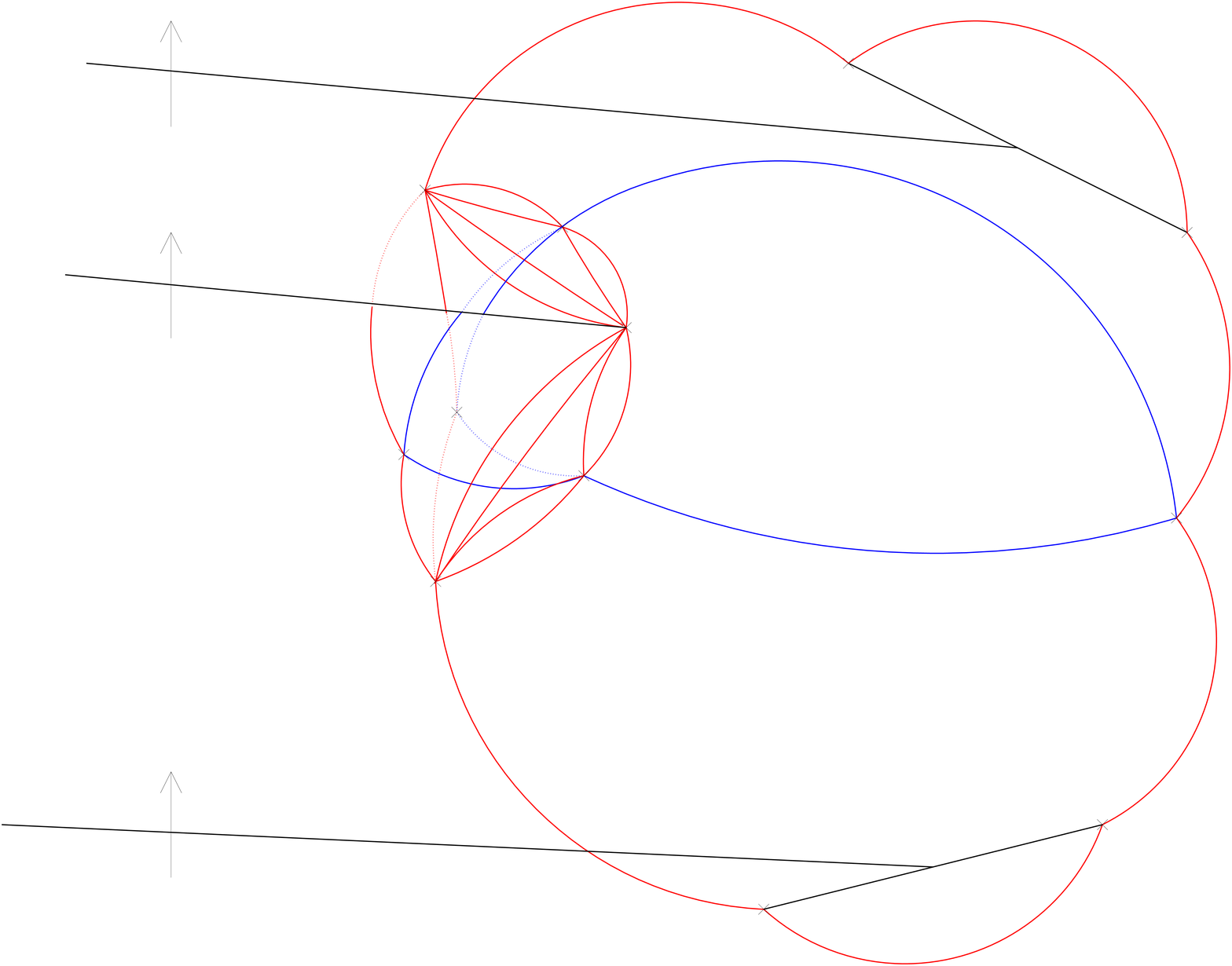}
\put(-100,100){double core} \put(-100,60){single core}
\put(-125,162){single core} \put(-202,146){$N_{1}$}
\put(-202,105){$[\lambda_{3}]^{N_{f}}$} \put(-202,5){$N_{2}$}
}\caption{When a quark singularity intersects a double core. The
two groups of three leaf-shaped curves signal the boundary of
stability for the states that have come through the cut in either
direction, as for the single core.}\label{dcorenf3b}}

\section{Flows Between the Examples}

As a further consistency check, as well as exploring an important
part of the parameter space, we should consider how, at least for
large $R/\Lambda$ the models with a greater number of flavours
flow into those with less as we take some of the $|m_{i}|$ to
infinity, causing them to decouple from the effective theory. For
simplicity we shall consider all the masses we take to infinity to
be equal, and also all those left behind equal.

If we begin with both sets of equal masses the same, we have a
quark singularity of order $N_{f}$. It is just the $N_{f}$-fold
product $[\lambda_{3}^{a}][\lambda_{3}^{b}]...$. If we give extra
(equal, here) mass to $N_{1}$ of them, then this singularity
simply splits into the appropriate combinations of $N_{1}$ and
$N_{f}-N_{1}$ of them. The singularity corresponding to the quarks
with larger $|m_{i}|$ will be closer to the origin, as this will
shrink to a point and disappear before the other one. All the
curves which touched the singularity before it split had a
$\lambda_{3}^{i}$ as a constituent of the marginal bound state.
These curves will bifurcate, diverging from superposition to
follow whichever of the two singularities contains the
$\lambda_{3}$ with their s-charge label. At a fixed value of $R$
and $\Lambda$, $R\gg \Lambda$, increasing the bigger mass will see
the corresponding quark singularity move ever closer to the
origin. At a certain mass, it will have to pass through the core.
This marks the transition from a space which looks largely like
the $N_{f}$ moduli space, with small areas different, to one which
looks largely like the $N_{f}-N_{1}$ space with a few isolated
pockets of $N_{f}$-ness remaining, which shrink away as the mass
increases further. As can be seen from fig.\
\hspace{-3pt}(\ref{splitqsing}), before we reach that stage, the
states which exist between the singularities transform nicely
across the monodromy of the remaining fixed mass quarks.

\FIGURE[h]{ \centering
\makebox[11cm]{\includegraphics[width=11cm]{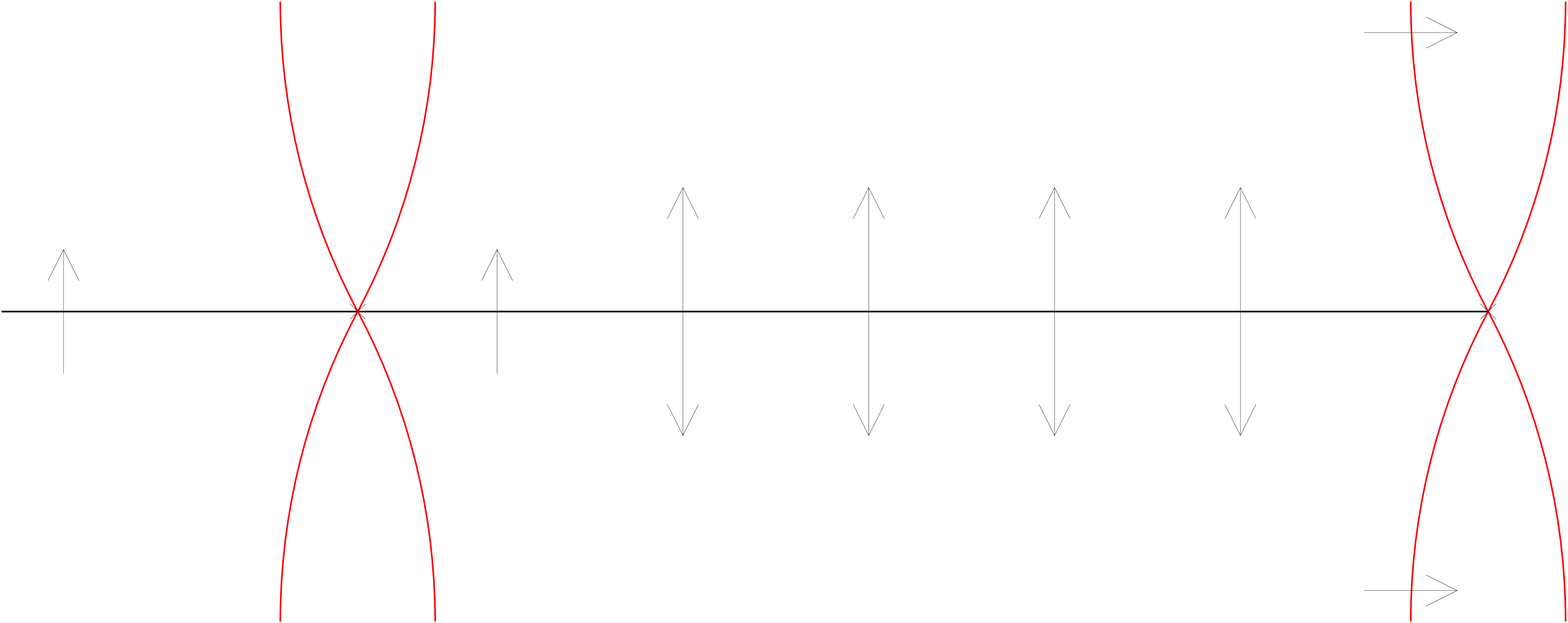}
\put(-330,39){$[\lambda_{3}^{a}][\lambda_{3}^{b}][\lambda_{3}^{c}]$}
\put(-222,39){$[\lambda_{3}^{c}]$}
\put(-180,92){$Q_{2}$} \put(-143,92){$\bar{Q}_{2}^{c}$}
\put(-106,92){$Q_{1}$} \put(-68,92){$\underline{Q}_{1}^{c}$}
\put(-182,26){$\underline{Q}_{2}^{c}$} \put(-145,26){$Q_{2}$}
\put(-108,26){$\bar{Q}_{1}^{c}$} \put(-70,26){$Q_{1}$}
\put(-55,115){$\underline{Q}_{1}^{c}$} \put(-20,115){$Q_{1}$}
\put(-16,101){$\lambda^{c}$}
\put(-55,3){$\underline{Q}_{2}^{c}$} \put(-20,3){$Q_{2}$}
\put(-16,16){$\lambda^{c}$}
\put(-3,70){$Q_{1}$} \put(-3,55){$Q_{2}\,\,\quuu_{3\mbox{-}}$}
\put(-3,40){$Q_{3\mbox{-}}$}
}\caption{The reduced spectrum left between the two quark
singularities consists of only those states that would be present
for the lower, right-hand set, plus $\alpha_{3}$ states invariant
under all the $[\lambda_{3}]$'s. Here we pick $N_{f}=3$ with two
large masses.}\label{splitqsing}}

As the divergent mass becomes larger, so too will the areas bound
by curves which pass through its quark singularity. Inside these,
the space carries only the smaller $N_{f}-N_{1}$ spectrum. If we
extend our usual radial section to include the diametrically
opposite half-plane too, as in fig.\
\hspace{-3pt}(\ref{sweeptorus}), we see that eventually the curves
on both sides pass through infinity on our stereographic
projection of the three-sphere. Continuing further, they restrict
more and more of the space to be $N_{f}-N_{1}$, and act as the
boundary confining the $N_{f}$ region inside a central sphere and
an annulus each touching the circle that is the heavier quark
singularity. After this circle touches the core, the annulus
detaches and then both it and the sphere shrink away to nothing at
a finite value of $|m|$ dependent on $R$ and $\Lambda$. Any states
of the highest $N_{f}$ bars, formerly inside both sets of curves,
will decouple as their mass tends to infinity.

\FIGURE[h]{ \centering
\makebox[13cm]{\includegraphics[width=14cm]{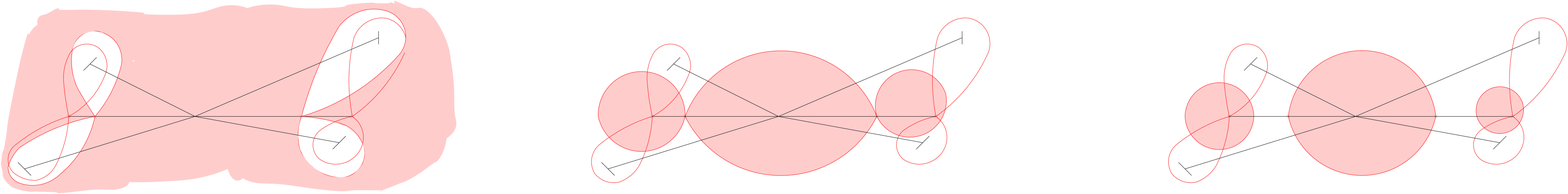}
}\caption{The shaded region represents the area where $\alpha_{1}$
and $\alpha_{2}$ magnetically charged dyons form bound states with
the quarks which have the bare mass we take to infinity. There are
3 regimes: first the curves grow, then, when the quark singularity
crosses the core, they merge at infinity, recombining to bound two
regions. Finally, these split and shrink away.}\label{sweeptorus}}

The one flavour case is not so special if we now consider how, for
$R\gg \Lambda$, the number of twists that the core undergoes
changes by $N_{1}$ half-twists. As we increase the mass, in all
cases in this regime, the first time the singularity surface
intersects is when the $N_{1}$-fold quark singularity crosses the
trefoliate curve with $N_{f}$ half-twists. This singularity is a
separate curve from the trefoil if and only if $N_{1}\geq 2$. This
could be seen to be the consequence of a Higgs branch being
incident at this singularity, or just that such a multiple
singularity cannot be equal to any of the single ones of the
trefoliate curve. In any case, if we always let $N_{1}=1$,
decoupling one flavour at a time, then the quark singularity must
always merge with the other one. If we do this many times, we
could alternatively have decoupled all at once. Thus we can tell
that we must see `demerging' too - the circle must reform on the
other side of the trefoliate curve. Each merger-demerger has the
effect of reducing the number of half-twists of the trefoliate
curve by one. We believe this may occur as in fig.\
\hspace{-3pt}(\ref{mergdemerg}). When a double, or stronger, quark
singularity passes through a core, we believe the two branches of
the trefoliate curve simply intersect, before, during or after, to
unwind the required number of times.

\FIGURE[h]{ \centering
\makebox[13cm]{\includegraphics[width=14cm]{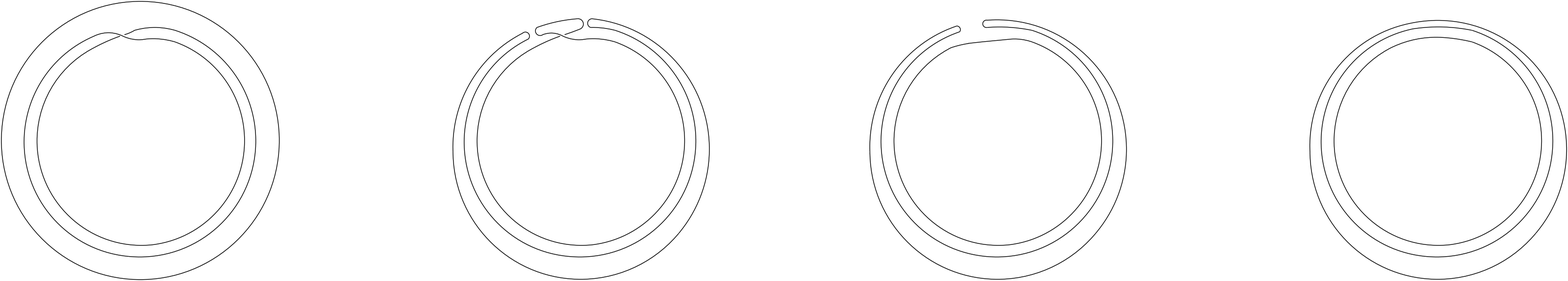}
}\caption{The outer curve represents the quark singularity, the
inner one the trefoliate curve. In the second figure they merge in
two places. The curve loses a half-twist and then they
demerge.}\label{mergdemerg}}

The interested reader may have realized that in the other sectors
of the moduli space, where $\alpha_{3}$ is not the non-simple
root, the monodromies for the third root are different, and
moreover depend on the number of flavours. In each case we have
not $N_{3-}=[Q_{3-}^{n}][Q_{3-}^{n-1}]$, but the most upper-barred
equivalent, {\it e.g.\ }\hspace{-3pt}
$\raisebox{2pt}{$\bar{\hspace{9.5pt}}$}\hspace{-9pt}\raisebox{3.5pt}{$\bar{\hspace{9.5pt}}$}\hspace{-11pt}
\bar{N}_{3-}=[\quuu_{3-}^{n}][ \quuu_{3-}^{n-1}]$. The only way to
enforce a change at the transition is to sweep around this third
of the trefoil with the departing quark monodromy, in much the
same way as in \cite{bf3} for SU$(2)$. We show this in
diagrammatic form in fig.\ \hspace{-3pt}(\ref{sweepmonod}). Our
ability to do this is based on the group relation
\[ \bar{N}_{3-}=[\lambda_{2}]^{-1}N_{3-}[\lambda_{2}]
=[\lambda_{1}]N_{3-}[\lambda_{1}]^{-1} \] which also holds if we
put an extra bar on all the $N_{3}$'s, any number of times.

\FIGURE[h]{ \centering
\makebox[13cm]{\includegraphics[width=14cm,height=5.5cm]{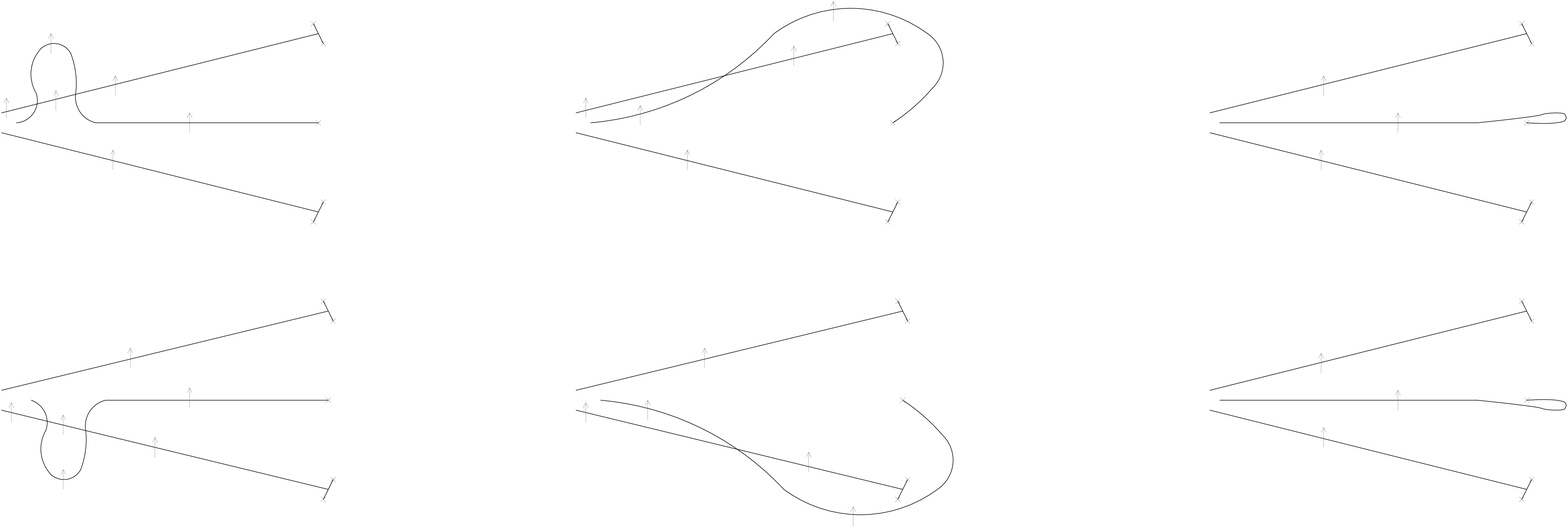}
\put(-401,129){\scriptsize$N_{3}$}
\put(-390,131){\scriptsize$N_{3}$}
\put(-374,136){\scriptsize$\bar{N}_{3}$}
\put(-392,150){\scriptsize$[\lambda_{2}]$}
\put(-357,125){\scriptsize$[\lambda_{2}]$}
\put(-374,100){\scriptsize$N_{1}$}
\put(-400,25){\scriptsize$N_{3}$}
\put(-388,21){\scriptsize$N_{3}$}
\put(-364,13){\scriptsize$\bar{N}_{3}$}
\put(-390,5){\scriptsize$[\lambda_{1}]$}
\put(-357,29){\scriptsize$[\lambda_{1}]$}
\put(-369,55){\scriptsize$N_{2}$}
\put(-198,24){\scriptsize$N_{3}$}
\put(-255,23){\scriptsize$N_{3}$}
\put(-188,-7){\scriptsize$[\lambda_{1}]$}
\put(-239,38){\scriptsize$[\hspace{-0.4pt}\lambda_{2}]$}
\put(-223,55){\scriptsize$N_{2}$}
\put(-201,131){\scriptsize$N_{3}$}
\put(-254,129){\scriptsize$N_{3}$}
\put(-194,159){\scriptsize$[\lambda_{2}]$}
\put(-239,116){\scriptsize$[\hspace{-0.4pt}\lambda_{1}]$}
\put(-228,100){\scriptsize$N_{1}$}
\put(-65,137){\scriptsize$N_{3}$}
\put(-50,127){\scriptsize$[\lambda_{1}]$}
\put(-67,101){\scriptsize$N_{1}$}
\put(-66,53){\scriptsize$N_{2}$}
\put(-50,28){\scriptsize$[\lambda_{2}]$}
\put(-67,17){\scriptsize$N_{3}$}
}\caption{How the cut of the singularity of quarks with bare mass
tending to infinity loops around the third arm (to which we have
paid little attention so far) before shrinking. $R$ increases to
the right.}\label{sweepmonod}}

The introduction of bare masses changes the position of all the
multiple points of the singularity curve. The first intersections,
just described, move outwards such that when we reach the regime
$m\gg \Lambda$, $m={\rm Max}\, |m_{i}|$, they occur at scale
$R\sim m$. All the others vary with smaller powers of $m$, none
remaining fixed. One could think of this as marking a steady
transition from dependence on $\Lambda_{N_{f}}$ through to
$\Lambda_{N_{f}-N_{1}}=(m_{1}\dots
m_{N_{1}}\Lambda_{N_{f}}^{6-N_{f}})^{1/(6-N_{f}+N_{1})}$. We
cannot describe in detail the tangled webs which ensue, but merely
point out that when we add masses, the multiple intersections tend
to split, and all the discrete symmetry of the model is lost as
the positions of the intersections move towards those of the
theory with a lower number of flavours. Upon taking these masses
to infinity, these intersection points merge with others to form
higher order intersections again.

\section{Conclusions}

We have thus described qualitatively, in some, but not complete
detail, a consistent model of low-energy effective ${\cal N}=2$
supersymmetric gauge theory for SU$(3)$ with up to 5 flavours of
hypermultiplets in the fundamental. We have concentrated on the
regime where $R\gg \Lambda_{N_{f}}$ and listed the BPS spectra
which occur here. Just as we found for the pure case, there are
mutually disjoint areas in the weak-coupling limit carrying
different BPS spectra. In each of these there exist
$3=N_{c}(N_{c}-1)/2$ infinite towers of stable BPS dyons. We have
further $3N_{f}$ quarks, and $2^{N_{f}}$ possible bound states of
these with each element of each of the three towers. In a
particular region, it may instead be energetically favourable for
one of the towers to bind with antiquarks.

There are always single cores --- areas where one of the remnant
U$(1)$'s is strongly coupled. There is also an area where some of
the quarks (one of the elements of the colour triplet for each
flavour) become light. There are then regions which strictly
include the core, and some of the latter area, in which most of
these dyon-quark bound states either lose stability, or,
interestingly, bind further with more quarks, until they have one
of each flavour and so become a flavour singlet. This state of
affairs persists in the core, where the bound states occur only
with the reduced number of those dyons which exist there in the
pure model.

We have additional, and rather complicated singularity structure
to that of the pure case. For fixed large $R/\Lambda$ we see a
`trefoliate' curve which twists about a trefoil, and a circular
`quark singularity' curve for each flavour, coincident for those
flavours which possess the same bare mass. At least for when we
have some large masses, we can be sure that these two intersect
before other self-intersections complicate the picture. Then the
quark singularity must pass through the single core. Double cores
 also exist, at the intersection of single ones. We are then led
to the question of whether the quark singularity passes through
these.

We have also tweaked the model we presented for the pure case
\cite{me}. The addition of curves upon which higher spin
multiplets become unstable accommodates the conventional weak
coupling spectrum, without needing to invoke the existence of as
many of these states co-existing with simple root magnetic charge.
Despite impressive machinery \cite{piljin2}, the existence of
these states has yet to be disproved, so we offer this model as an
(more likely) alternative rather than a correction. Within the
double core, the use of an extra branch cut does seem a
correction, allowing us a smaller, more symmetric BPS spectrum,
which transforms more clearly as the quantum monodromies pass
through each other. We think a model along the lines of \cite{me}
would still be possible, but it becomes almost intractably messy.

Based on \cite{hh}, in \cite{nick1,dnt} Dorey {\it et al.\
}\hspace{-3pt} argued that at a particular point of the
SU$(N_{c})$ moduli space with matter the massive BPS spectrum is
equivalent to a particular two-dimensional $(2,2)$-SUSY gauge
theory with U$(N_{c})$ flavour symmetry. In one regime, the
corresponding point in Seiberg-Witten is at weak-coupling.
However, the 2D theory has a limit with a dual description and a
different, smaller, calculable BPS spectrum corresponding to when
our point is at strong coupling with a reduced set of stable
states. There is scope, then, for our predictions to be matched
against each other.

Thus far we have given the impression that the quark singularities
are coincident if and only if the bare masses of all the flavours
are the same. This is not quite true. When $N_{f}\geq N_{c}$ there
is one other possibility, which occurs when the auxiliary curve
degenerates into two equal factors. For SU$(3)$ with three
flavours this occurs when the masses $m_{i}$ are the three roots
of $m_{i}^{3}-u m_{i}+v=0$. Necessarily they sum to zero. So for
each $u,\, v$ we could find masses such that all the quark
singularities (or whatever they have become at strong coupling)
are coincident, and the auxiliary curve a square. Alternatively,
for each set of $m_{i}$ summing to zero, we could find the $u,\,
v$ coordinates of where such a singularity would occur. For four
and five flavours, this class falls into the set of occurrences
for which three singularities are coincident (the others
coincident too as special cases). The Galois group of the defining
equations treats three, say $m_{1},\, m_{2},\, m_{3}$, differently
to the other one or two. The theme linking these events is that in
each case these $^{N_{\hspace{-1pt}f}}\hspace{-1pt}C_{3}$ order 3
singularities are also the root of a baryonic Higgs branch.

Although we offer no proof of the fact, we expect the theories of
three equal masses and of those at the root of this baryonic
branch to be very similar. For SU$(2)$, where the relations for
two flavours to be at this root are $m_{1}=-m_{2}=m$, and
$u=m^{2}-\Lambda^{2}/8$, as opposed to $m_{1}=m_{2}$ and
$u=m^{2}+\Lambda^{2}/8$. If, for the former, we let $m\rightarrow
im$, we get the mirror image moduli space, the only difference we
find between whether $m_{2}=\pm m_{1}$ is that of the natural
assignment of s-charges. For SU$(3)$, we find only a slight
discrepancy: the values of $u$ and $v$ for three equal masses are
given implicitly as those which solve $m^{3}-u
m+v=k(\Lambda_{N_{f}})$ where $k(\Lambda_{3})=\Lambda_{3}^{3}/4$,
$k(\Lambda_{4})=m\Lambda_{4}^{2}/2$ and
$k(\Lambda_{5})=m^{2}\Lambda_{5}/4$. The authors of \cite{dnt}
argue that the baryonic Higgs branch has a strong-coupling region
with only a finite spectrum. For us this implies that for three or
more flavours, the quark singularity must pass through a double
core. We cannot definitively say whether or not this is true, but
it seems a perfectly reasonable assumption. Given this, we agree
on both weak and strong coupling spectra, with possible
differences in that in almost all cases we have the intermediately
barred states only at most marginally stable in the neighbourhood
of the singularity, and also that we have included regions in this
neighbourhood where bound states with the antiquarks, rather than
with the quarks exist. Thus modulo these slight discrepancies, we
find ourselves to be in broad agreement, lending weight to both
arguments.

Finally we should state that it would be fascinating to extend
this analysis to the case of six flavours where other studies
exist \cite{buc1,buc2}. There are also string web constructions
and other methods involving the solution of geodesic equations,
such as \cite{bs, war, sethi,berg,henyi} which could in principle
reproduce these results. We would like, eventually, to write $a$
and $a_{D}$ as ratios of two-variable theta functions and then to
plot the positions of the CMS's exactly. This is very much more
involved than for SU$(2)$ where the underlying mathematical
structure is well documented.

\acknowledgments This work was supported by the European Research
and Training Network ``Superstring Theory" (HPRN-CT-2000-00122).

We would like to thank David Olive, Adam Ritz, Dave Tong and
Piljin Yi for useful and stimulating discussions.

\appendix
\section{Singularity Curves}

We choose to use the auxiliary curves of \cite{hanoz}, in Table
\ref{tab1} we list first the massless cases, then those of general
mass.

\TABLE[b]{\begin{tabular}{|c|l|}\hline & \\[-10pt] $N_{f}$ & $y^{2}=$ \\[3pt]
\hline & \\[-10pt]
$0$ & $(x^{3}-ux-v)^{2}-\Lambda^{6}_{0}$ \\
$1$ & $(x^{3}-ux-v)^{2}-\Lambda^{5}_{1}x$ \\
$2$ & $(x^{3}-ux-v)^{2}-\Lambda^{4}_{2}x^{2}$ \\
$3$ & $(x^{3}-ux-v+ \Lambda_{3}^{3}/4)^{2}-\Lambda^{3}_{3}x^{3}$ \\
$4$ & $(x^{3}-ux-v+\Lambda_{4}^{2}x/4)^{2}-\Lambda^{2}_{4}x^{4}$ \\
$5$ & $(x^{3}-ux-v+\Lambda_{5}x^{2}/4)^{2}-\Lambda_{5}x^{5}$
\\[3pt]
\hline & \\[-10pt]
$1m$ & $(x^{3}-ux-v)^{2}-\Lambda^{5}_{1}(x+m_{1})$ \\
$2m$ & $(x^{3}-ux-v)^{2}-\Lambda^{4}_{2}(x+m_{1})(x+m_{2})$ \\
$3m$ & $(x^{3}-ux-v+ \Lambda_{3}^{3}/4)^{2}-\Lambda^{3}_{3}(x+m_{1})(x+m_{2})(x+m_{3})$ \\
$4m$ & $(x^{3}\hspace{-1pt}-\hspace{-1pt}ux\hspace{-1pt}-\hspace{-1pt}v\hspace{-1pt}+\hspace{-1pt}\Lambda_{4}^{2}(x\hspace{-1pt}
+\hspace{-1pt}m_{1}\hspace{-1pt}+\hspace{-1pt}m_{2}\hspace{-1pt}+\hspace{-1pt}m_{3}\hspace{-1pt}+
\hspace{-1pt}m_{4})/4)^{2}\hspace{-1pt}-\hspace{-1pt}\Lambda^{2}_{4}(x\hspace{-1pt}+\hspace{-1pt}m_{1})
(x\hspace{-1pt}+\hspace{-1pt}m_{2})(x\hspace{-1pt}+\hspace{-1pt}m_{3})(x\hspace{-1pt}+\hspace{-1pt}m_{4})$ \\
$5m$ &
$(x^{3}\hspace{-1pt}-\hspace{-1pt}ux\hspace{-1pt}-\hspace{-1pt}v\hspace{-1pt}+\hspace{-1pt}\Lambda_{5}(x^{2}\hspace{-2pt}
+\hspace{-1pt}(m_{1}\hspace{-2pt}+\hspace{-1pt}m_{2}\hspace{-1pt}+\hspace{-1pt}m_{3}\hspace{-1pt}+\hspace{-1pt}m_{4}\hspace{-1pt}
+\hspace{-1pt}m_{5})x\hspace{-1pt}+\hspace{-1pt}(m_{1}m_{2}\hspace{-1pt}+\hspace{-1pt}m_{1}m_{3}\hspace{-1pt}+\hspace{-2pt}\dots
\hspace{-2pt}+\hspace{-1pt}m_{4}m_{5})) /4)^{2}$\\ &
$-\Lambda_{5}(x+m_{1})(x+m_{2})(x+m_{3})(x+m_{4})(x+m_{5})$ \\
\hline
\end{tabular}\caption{A list of the auxiliary elliptic curves for massless, then general masses}\label{tab1}}

The singularities of each curve are the set of $u$ and $v$ where
the elliptic curve degenerates, {\it i.e.\ }\hspace{-3pt} has some
roots in common. This can be written as the zero set of a
polynomial, the discriminant. In Table \ref{tab2} we write this
for massless flavours.

\TABLE{\begin{tabular}{|c|l|}\hline & \\[-10pt] $\hspace{-1pt}N_{\hspace{-1pt}f}\hspace{-1pt}$ &
Discriminant \\[3pt]
\hline & \\[-10pt]
$0$ &
$\Lambda^{\hspace{-1pt}18}(4u^{3}-27(v+\Lambda^{3})^{2})(4u^{3}-27(v-\Lambda^{3})^{2})$
\\
$1$ &
$\Lambda^{\hspace{-1pt}15}(3125\Lambda^{\hspace{-1pt}15}\hspace{-3pt}-\hspace{-1pt}256\Lambda^{5}u^{5}\hspace{-3pt}-\hspace{-1pt}
22500\Lambda^{10}uv\hspace{-1pt}+\hspace{-1pt}1024u^{6}v
\hspace{-1pt}+\hspace{-1pt}43200\Lambda^{5}u^{2}v^{2}\hspace{-2pt}-\hspace{-1pt}13824u^{3}v^{3}\hspace{-2pt}
+\hspace{-1pt}46656v^{5})$\\
$2$ & $-64\Lambda^{12}v^{2}
(4\Lambda^{6}\hspace{-1pt}+12\Lambda^{4}u+12\Lambda^{2}u^{2}\hspace{-1pt}+4u^{3}\hspace{-1pt}-27v^{2})
(4\Lambda^{6}\hspace{-1pt}-12\Lambda^{4}u+12\Lambda^{2}u^{2}\hspace{-1pt}-4u^{3}\hspace{-1pt}+27v^{2})$\\
$3$ & $\frac{1}{16}\Lambda^{9}(\Lambda^{3}-4v)^{3}
(729\Lambda^{6}u^{3}\hspace{-1pt}-\hspace{-1pt}256u^{6}\hspace{-1pt}-\hspace{-1pt}3888\Lambda^{3}u^{3}v\hspace{-1pt}+
\hspace{-1pt}3456u^{3}v^{2}\hspace{-1pt} +\hspace{-1pt}2916
\Lambda^{3}v^{3}\hspace{-1pt}-\hspace{-1pt}
11664v^{4})$ \\
$4$ & $4\Lambda^{6}v^{4}
(-\Lambda^{2}u+8\Lambda^{2}u^{2}-16u^{3}+2\Lambda^{3}v-72\Lambda
uv+108v^{2})$ \\ &
$(-\Lambda^{2}u+8\Lambda^{2}u^{2}-16u^{3}-2\Lambda^{3}v+72\Lambda
uv+108v^{2})$\\
$5$ & $4\Lambda^{3}v^{5}
(-\Lambda^{2}u^{5}+256u^{6}-2\Lambda^{3}u^{3}v+528\Lambda
u^{4}v-\Lambda^{4}uv^{2}+300\Lambda^{2}u^{2}v^{2}$\\ &
$-3456u^{3}v^{2} -4\Lambda^{3}v^{3}+4212\Lambda u
v^{3}+11664v^{4})$ \\[1pt] \hline
\end{tabular}\caption{The discriminants for massless flavours}\label{tab2}}

\TABLE[b]{
\begin{tabular}{c|c|c|c|c|c|c}$N_{f}$ & fac$^{\rm \underline{n}}$ & sing facs & class & $u$ &
$v$ & $R$ \\ \hline $0$ & $AB$ & $AB$ & $*M^{2}_{2}$ &
$3\Lambda^{2}/\sqrt[3]{4}$ & $0$ & $27$ \\
&&$A^{2},B^{2}$&$*M^{0}_{3}$&$0$&$\pm \Lambda^{2}$&$27$ \\
$1$&$A$&$A^{2}$&$M^{1}_{0}$&$5\Lambda^{2}\sqrt[5]{3/2^{12}}$
&$5\Lambda^{3}/\sqrt[5]{2^{13}3^{6}},\,(uv$
&$25\sqrt[5]{2^{-21}3^{8}}$\\
&&$A^{2}$&$M^{1}_{0}$&$-5\Lambda^{2}/\sqrt[5]{2^{8}3}$
&$-5\Lambda^{3}/\sqrt[5]{2^{2}3^{9}},\in\mathbb{R})$
&$25\sqrt[5]{2^{-14}3^{7}}$\\
$2$&$Q^{2}AB$&$Q^{2}A^{2},Q^{2}B^{2}$&$*M^{2}_{4}$&$\pm\Lambda^{2}$&$0$&$4$\\
&&$AB$&$M^{2}_{0}$&$\pm i \Lambda^{2}/\sqrt{3}$&$\sqrt{\pm
32i/81\sqrt{3}}$&$1/\sqrt{3}$\\
$3$&$Q^{3}A$&$Q^{3}A$&$M^{3}_{2}$&$0$&$\Lambda^{3}/4$&$27/16$\\
&&$Q^{3}A$&$M^{3}_{2}$&$-3\Lambda^{2}/\sqrt[3]{2^{8}}$&$\Lambda^{3}/4$&$3^{3}5/64$\\
&&$A^{3}$&$M^{3}_{0}$&$-3^{2}\Lambda^{2}/\sqrt[3]{2^{12}}$&$9\Lambda^{3}/32$&$3^{6}/2^{8}$\\
$4$&$Q^{4}AB$&$Q^{4}AB$&$M^{4}_{2}$&$0$&$0$&$0$\\
&&$Q^{4}AB$&$*M^{4}_{4}$&$\Lambda^{2}/4$&$0$&$1/16$\\
&&$AB$&$M^{4}_{0}$&$\Lambda^{2}/36$&$2\Lambda^{3}/\sqrt{3^{9}}$&$65/3^{6}2^{4}$\\
&&$A^{2},B^{2}$&$M^{4}_{0}$&$\Lambda^{2}/12$&$\mp \Lambda^{3}/27$&$17/27$\\
$5$&$Q^{5}A$&$Q^{5}A^{3}$&$M^{5}_{4}$&$0$&$0$&$0$\\
&&$Q^{5}A$&$M^{5}_{2}$&$\Lambda^{2}/2^{8}$&$0$&$1/2^{22}$\\
&&$A^{2}$&$M^{5}_{0}$&$\frac{39\pm
55\sqrt{33}}{2^{13}3^{2}}\Lambda^{2}$&$\frac{2879\mp
561\sqrt{33}}{2^{18}3^{3}}\Lambda^{3}$&$
\sim5.1\hspace{-2pt}\times\hspace{-2pt}10^{-7}$\\ &&&&&&$2.0\hspace{-2pt}\times\hspace{-2pt}10^{-5}$\\
&&$A^{2}$&$M^{5}_{0}$&$\Lambda^{2}/3^{5}$&$-\Lambda^{3}/2^{13}3^{2}$&$5045113/2^{16}3^{15}$\\
\end{tabular}\caption{Intersection points of the singularity curves for massless matter.}\label{tab3}}

The main features of the discriminants are how they factorize, and
their self-intersections, or multiple zeroes. In Tables \ref{tab3}
and $5$ we note the position and order of most of them (for equal
masses). We do not claim this to be comprehensive and exhaustive,
except for zero masses, but based purely on those that could
readily be open to solution by radicals using Mathematica and
expressible in a small space. The first column denotes which
example we are considering, {\it e.g.}, $4m2$ is four flavours,
two of which are massive with equal mass $m$. In the second, we
state the number of factors, by labelling each quark singularity
as $Q$, and each other factor with a capital letter $A, B, \dots$.
In the third column is a word where we include one occurrence of
each factor for each order of zero that factor has at the
singularity. The number of these gives the order of the
singularity, {\it i.e.\ }\hspace{-3pt} the first order of
derivatives where any one of which has a non-zero value at the
singularity. It is useful to distinguish these singularities by
the universality class of the superconformal field theory which
exists there, as defined in \cite{eguchi}, so we state this in
column four. We add a star if this is not of the trivial class 1
in \cite{eguchi}. Columns five and six state, where possible, the
($u,v$) positions of the singularity, and the last column, the
sixth power of the radius in units of $\Lambda$, {\it i.e.\
}\hspace{-3pt} $(4|u|^{3}+27|v|^{2})/\Lambda^{6}$. Note that where
square and cube roots, {\it etc.}, are given as coordinates, we
imply that there is one of these roots at each of the possible
values.

\TABLE{
\begin{tabular}{c|c|c|c|c|c}$N_{f}$ & fac$^{\rm \underline{n}}$ & sing facs & ord & $u$ &
$v$ \\ \hline
$2m2$&$Q^{2}AB$&$Q^{2}A,Q^{2}B$&$3$&$3m^{2}\pm\Lambda^{2}$&$m(2m^{2}\pm\Lambda^{2})$ \\
&&$Q^{2}A,Q^{2}B$&$3$&$(3m^{2}\pm
4\Lambda^{2})/4$&$m(\pm\Lambda^{2}-m^{2}/4)$ \\
&&$A^{2},B^{2}$&$2$&$\pm\Lambda^{2}$&$\pm\Lambda^{2}m$ \\
&&$AB$&$2$& 4 points & \\
$3m1$&$Q^{2}A$&$Q^{2}A$&$3$&$\sqrt{\Lambda^{3}m}$&$\Lambda^{3}/4$ \\
&&$Q^{2}A$&$3$&3 points&$\Lambda^{3}/4$ \\
$3m2$&$Q^{2}A$&$Q^{2}A$&$2$&$3m^{2}\hspace{-2pt}\pm\hspace{-2pt} i
\sqrt{\Lambda^{3}m}$&$\frac{\Lambda^{3}+8m^{3}+4mi\sqrt{\Lambda^{3}m}}{4}$ \\
&&$Q^{2}A$&$3$& 3 points& \\
$3m3$&$Q^{3}A$&$Q^{3}A$&$4$&$3m^{2}$&$(\Lambda^{3}+8m^{3})/4$ \\
&&$Q^{3}A$&$4$& 3 points & \\
$4m1$&$Q^{3}A$&$Q^{3}A$&$4$&$\Lambda^{2}/4$&$\Lambda^{2}m/4$ \\
&&$Q^{3}A$&$4$& 4 points &$\Lambda^{2}m/4$ \\
$4m2$&$Q_{1}^{2}Q_{2}^{2}AB$&$Q_{1}^{2}Q_{2}^{2}$&$4$&$(\Lambda^{2}+4m^{2})/4$&$\Lambda^{2}m/2$
\\
&&$Q_{1}^{2}A,Q_{1}^{2}B$&$3$&$(\pm3m^{2}-2\Lambda
m)4$&$\hspace{-4pt}\frac{\Lambda m(\Lambda \mbox{-}2m)+m^{3}(\pm 3 \mbox{-}4)}{4}\hspace{-4pt}$ \\
&&$Q_{1}^{2}A,Q_{1}^{2}B$&$3$&$(\Lambda^{2}\pm4\Lambda
m+12m^{2})/4$&$\frac{2\Lambda^{2}m\pm\Lambda m^{2}+8m^{3}}{4}$ \\
&&$Q_{2}^{2}A,Q_{2}^{2}B$&$3$&$(\Lambda^{2}\pm4\Lambda m)/4 $&$\Lambda^{2} m/2$ \\
&&$Q_{2}^{2}A,Q_{2}^{2}B$&$3$&$\pm\Lambda m $&$\Lambda^{2} m/2$ \\
&&$A^{2},B^{2}$&$2$&$-(\Lambda^{2}\pm12\Lambda m)/4$&$(27m\pm2\Lambda)/54$ \\
&&$AB$&$2$& 5 points & \\
$4m3$&$Q^{3}A$&$Q^{3}A$&$4$&$(\Lambda^{2}+12m^{2})/4$&$\frac{27m\pm2\Lambda}{54}\Lambda^{2}$ \\
&&$Q^{3}A$&$4$& 4 points & \\
$4m4$&$Q^{4}AB$&$Q^{4}AB$&$6$&$(\Lambda^{2}+12m^{2})/4$&$m(\Lambda^{2}+2m^{2})$ \\
&&$Q^{4}A,Q^{4}B$&$5$&$\frac{3m}{4}(2\Lambda\pm m)$&$\frac{3\Lambda^{2}m+6\Lambda m^{2}-(4\mp3)}{4}$ \\
&&$AB$&$2$&$(\Lambda^{2}+12m^{2})/4$&$\frac{\Lambda^{2}m(\Lambda+7m)(2\Lambda+19m)}{12\Lambda^{2}+27\Lambda m+27m^{2}}$ \\
&&$A^{2},B^{2}$&$2$&$-(\pm\Lambda^{2}+24\Lambda
m)/12$&$\frac{\Lambda}{27}(\Lambda^{2}\hspace{-2pt}\pm\hspace{-2pt}27\Lambda
m\hspace{-2pt}-\hspace{-2pt}27m^{2})$ \\
&&$AB$&$2$& 3points & \\
$5m1$&$Q^{4}A$&$Q^{4}A^{2}$&$6$&$\Lambda m/4$&$0$ \\
&&$Q^{4}A^{2}$&$6$&$0$&$0$ \\
&&$Q^{4}A$&$5$&$(\Lambda +16m)^{2}/2^{8}$&$0$ \\
$5m2$&$Q_{1}^{2}Q_{2}^{3}A$&$Q_{1}^{2}A$&$3$&$3m^{2}\pm i \sqrt{\Lambda m^{3}}$&$2m^{2}\Lambda\pm i m\sqrt{m^{3}\Lambda}$ \\
&&$Q_{1}^{2}$&$3$&$\frac{\Lambda^{2}+48\Lambda m+384m^{2}\pm\sqrt{\Lambda(\Lambda+32m)^{3}}}{2^{9}}$&$m(u-m^{2})$ \\
&&$Q_{2}^{3}A$&$5$&$m^{2}+\Lambda m/4$&$\Lambda m^{2}/4$ \\
&&$Q_{2}^{3}A$&$4$&$\Lambda m/2$&$\Lambda m^{2}/4$ \\
&&$Q_{2}^{3}A$&$4$& 3 points &$\Lambda m^{2}/4$ \\
\end{tabular}\label{tab4}}
\TABLE{
\begin{tabular}{c|c|c|c|c|c}$N_{f}$ & fac$^{\rm \underline{n}}$ & sing facs & ord & $u$ &
$v$ \\ \hline
$5m3$&$Q_{1}^{3}Q_{2}^{2}A$&$Q_{1}^{3}A$&$4$&$(\Lambda m+12m^{2})/4$&$(2m^{2}\Lambda+8m^{3})/4$ \\
&&$Q_{1}^{3}A$&$4$& 3 points & \\
&&$Q_{1}^{3}Q_{2}^{2}$&$5$&$(11\Lambda m+4m^{2})/4$&$3\Lambda m^{2}/4$ \\
&&$Q_{2}^{2}A$&$3$&$(3\Lambda m\pm4\sqrt{\Lambda m^{3}})/4$&$3\Lambda m^{2}/4$ \\
&&$Q_{2}^{2}A$&$3$&$\frac{\Lambda(\Lambda+96m)\pm\sqrt{(\Lambda+96m)^{3}}}{2^{9}}$&$3\Lambda m^{2}/4$ \\
$5m4$&$Q^{4}A$&$Q^{4}A$&$5$&$(\Lambda m+6m^{2})/2$&$(5\Lambda m^{2}+8m^{3})/4$ \\
&&$Q^{4}A$&$5$& 3 points & \\
$5m5$&$Q^{5}A$&$Q^{5}A$&$6$&$(3\Lambda m+12m^{2})/4$&$(9\Lambda m^{2}+8m^{3})/4$ \\
&&$Q^{5}A$&$6$&$\frac{\Lambda^{2}+144\Lambda m+384m^{2}\pm\sqrt{\Lambda(\Lambda+96m)^{3}}}{2^{9}}$&
$(3\Lambda m\hspace{-2pt}-\hspace{-2pt}2m^{3}\hspace{-2pt}+\hspace{-2pt}2mu)/2$ \\
\end{tabular}\label{tab5}}

\end{document}